%% file: acmlarge.tex
\documentclass[acmlarge]{acmart}
\input{contents/heading}

\AtBeginDocument{%
  \providecommand\BibTeX{{%
    \normalfont B\kern-0.5em{\scshape i\kern-0.25em b}\kern-0.8em\TeX}}}

\setcopyright{acmcopyright}
\acmJournal{IMWUT}
\acmYear{2021} \acmVolume{5} \acmNumber{3} \acmArticle{143} \acmMonth{9} \acmPrice{15.00}\acmDOI{10.1145/3478120}

\begin{document}

%%
%% The "title" command has an optional parameter,
%% allowing the author to define a "short title" to be used in page headers.
\title{SyncUp: Vision-based Practice Support for Synchronized Dancing}

%%
%% The "author" command and its associated commands are used to define
%% the authors and their affiliations.
%% Of note is the shared affiliation of the first two authors, and the
%% "authornote" and "authornotemark" commands
%% used to denote shared contribution to the research.

\author{Zhongyi Zhou}
\affiliation{
  \institution{Interactive Intelligent Systems Lab., The University of Tokyo}
  \city{Tokyo}
  \country{Japan}
}
\email{zhongyi@iis-lab.org}
\author{Anran Xu}
\affiliation{
  \institution{Interactive Intelligent Systems Lab., The University of Tokyo}
  \city{Tokyo}
  \country{Japan}
}
\email{anran@iis-lab.org}
\author{Koji Yatani}
\affiliation{
  \institution{Interactive Intelligent Systems Lab., The University of Tokyo}
  \city{Tokyo}
  \country{Japan}
}
\email{koji@iis-lab.org}
%%
%% By default, the full list of authors will be used in the page
%% headers. Often, this list is too long, and will overlap
%% other information printed in the page headers. This command allows
%% the author to define a more concise list
%% of authors' names for this purpose.
\renewcommand{\shortauthors}{Zhou et al.}

%%
%% The abstract is a short summary of the work to be presented in the
%% article.
\begin{abstract}
The beauty of synchronized dancing lies in the synchronization of body movements among multiple dancers. 
While dancers utilize camera recording\revise{}{s} for their practice, standard video interfaces do not efficiently support their activities of identifying segments where they are not well synchronized.
This thus fails to close a tight loop of an iterative practice process (i.e., capturing a practice, reviewing the video, and practicing again).
We present SyncUp, a system \revise{providing}{that provides} multiple interactive visualizations to support the practice of synchronized dancing \revise{to}{and} liberate users from manual inspection of recorded practice videos. 
By analyzing videos uploaded by users, SyncUp quantifies two aspects of synchronization in dancing: pose similarity among multiple dancers and temporal alignment of their movements. 
The system then highlights which body parts and which portions of the dance routine require further practice to achieve better synchronization. 
The results of our system evaluations show that our pose similarity estimation and temporal alignment predictions were correlated \revise{}{well }with human ratings\revise{ well}{}.
Participants in our qualitative user evaluation expressed the benefits and its potential use of SyncUp, confirming that it would enable quick iterative practice.
\end{abstract}

%%
%% The code below is generated by the tool at http://dl.acm.org/ccs.cfm.
%% Please copy and paste the code instead of the example below.
%%


%%
%% Keywords. The author(s) should pick words that accurately describe
%% the work being presented. Separate the keywords with commas.

%%
%% This command processes the author and affiliation and title
%% information and builds the first part of the formatted document.
\maketitle
\input{contents/intro}
\input{contents/related}
\input{contents/formative}
\input{contents/interface}
\input{contents/method}

\input{contents/eval}
\input{contents/userstudy}
\input{contents/conclusion}

%%
%% The next two lines define the bibliography style to be used, and
%% the bibliography file.
\bibliographystyle{ACM-Reference-Format}
\bibliography{reference}

\input{contents/appendix}

\end{document}

%% file: contents/heading.tex
\usepackage{xspace}
\usepackage{bbm}
\usepackage{amsmath}
\usepackage{svg}
\usepackage[normalem]{ulem}
\usepackage{xcolor,cancel}
\usepackage{mathtools}
\usepackage{dsfont}
\usepackage{caption}
\usepackage{subcaption}
\usepackage{eucal}

\usepackage{array}

\makeatletter
\def\Hline{
  \noalign{\ifnum0=`}\fi\hrule \@height 4.\arrayrulewidth \futurelet
  \reserved@a\@xhline}
\makeatother

\def\systemname{SyncUp\xspace}
\def\pgram{\textit{posegram}\xspace}
\def\pgrams{\textit{posegrams}\xspace}

\usepackage{amsmath}
\usepackage{graphicx}
\usepackage{here}
\PassOptionsToPackage{hyphens}{url}
\usepackage{hyperref} %% for hypertext
\usepackage{url} %これでURLがいい感じに折り返される
\usepackage{textcomp} %% for registared mark　→®
\usepackage{multicol}
\usepackage{subcaption}
\usepackage{comment}
\usepackage{booktabs,dcolumn}
\usepackage{multirow}
\usepackage{threeparttable}
\usepackage{siunitx} %% SI単位系の出力
\usepackage{fancyhdr}			% for use Pagestyle
\usepackage{totpages}
\usepackage{arydshln}
\usepackage{enumitem}
\usepackage{svg}
\usepackage{rotating}
\renewenvironment{quote}[1][0.04\linewidth]
  {\list{}{\leftmargin=#1\rightmargin=#1}\item\relax}{\endlist}
\newcommand{\myquote}[2]
{
\begin{quote}
\textit{#1} [#2]
\end{quote}
}

\usepackage{color}
\definecolor{orange}{RGB}{255,127,0}
\definecolor{darkgreen}{RGB}{0, 146, 0}
\definecolor{violet}{RGB}{148,0,211}

\newcolumntype{L}[1]{>{\raggedright\let\newline\\\arraybackslash\hspace{0pt}}m{#1}}
\newcolumntype{C}[1]{>{\centering\let\newline\\\arraybackslash\hspace{0pt}}m{#1}}
\newcolumntype{R}[1]{>{\raggedleft\let\newline\\\arraybackslash\hspace{0pt}}m{#1}}

\newif\ifCOMMENTS
\COMMENTSfalse
\ifCOMMENTS

\newcommand{\revise}[2]{\textcolor{red}{\sout{#1}#2}}

\else

\newcommand{\revise}[2]{#2}

\fi

\newcommand{\iconvideogreen}{\includegraphics[scale=0.15]{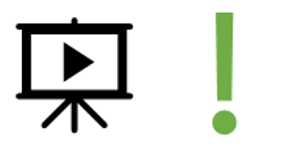}}%
\newcommand{\iconvideoblue}{\includegraphics[scale=0.15]{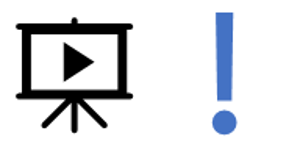}}%
\newcommand{\iconvideoorange}{\includegraphics[scale=0.15]{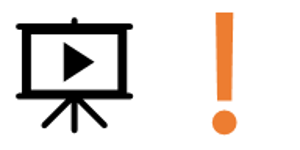}}%

\newcommand{\iconvideoblack}{\includegraphics[scale=0.15]{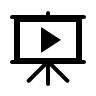}}%
\newcommand{\iconvideored}{\includegraphics[scale=0.15]{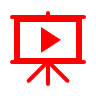}}%
\makeatletter
\def\Hline{
  \noalign{\ifnum0=`}\fi\hrule \@height 4.\arrayrulewidth \futurelet
   \reserved@a\@xhline}
\makeatother

%% file: contents/intro.tex
\begin{figure}
% \vspace{-1.5cm}
\centering
% \includesvg[width=\linewidth]{figures/IMWUT21_overview_TA.svg}
\includegraphics[width=\linewidth]{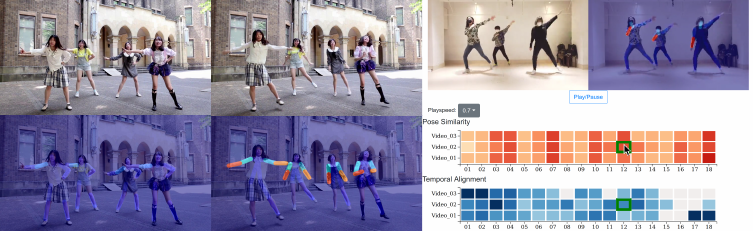}
\captionof{figure}{Left: SyncUp heatmap overlays on dancers highlight which body parts break synchronization in their poses. Colors closer to red and blue represent \revise{a }{}higher or lower degree\revise{}{s} of position discrepancy of each body part, respectively. Right: The SyncUp \revise{W}{w}eb interface. After dancers upload their practice videos, the system quantifies the degree of synchronization of their dance\revise{,}{} and offers feedback \revise{with}{using} three visualizations (\revise{}{i.e., }heatmap overlays on dancers' poses and two 1D heatmaps). This helps users quickly look for and replay a portion of their dances where there is an issue in synchronization.}
\label{fig:overview}
\end{figure}

\section{Introduction}

Synchronized dancing consists of a series of synchronized poses and/or temporally\revise{-}{ }aligned movements by multiple dancers\revise{,}{ that} creat\revise{ing}{e} visual aesthetics.
Many amateur dancers engage in such dances by replicating choreographies performed by popular idols and anime characters. 
They also share their performances \revise{on}{via} online social network services, often generating a huge popularity spike. 
It \revise{thus becomes}{is thus} an increasingly common practice for amateur dancers to record their dance performances.

Besides dissemination on social networks, dancers also use these recordings for their practice.
They use cameras on smartphones or tablets to capture their practices and share \revise{them}{the recordings} with other members to review \revise{which}{the} portion\revise{}{s} of their dance \revise{needs}{that require} improvements (e.g., moments where poses of the dancers are different or the timing of their motions are not aligned).
This is an important activity because it offers a good common ground \revise{among}{for} dancers \revise{on}{to understand} what \revise{}{needs }to be improved.
However, \revise{a}{the most} common approach for reviewing videos \revise{is still to use}{uses} a na\"{i}ve video player with simple navigation functionalities (e.g., \revise{a player built in}{on the} smartphone\revise{s}{}).
Even experienced dancers \revise{may need}{require} time and effort to identify moments where dancers are not well\revise{ }{-}synchronized.
Thus, dancers typically review their recordings after they finish all practices.
As a result, practices and their review activities are isolated, which can lead to less satisfactory user experience.

Existing research has examined interactive systems to support dance practices by recognizing dancers' poses and offering correction feedback~\cite{Anderson2013youmove, El2018Conceptual}.
These systems typically \revise{assume}{require} specialized hardware and infrastructure, making them less accessible to a general user population~\cite{Drobny2009Saltate, Anderson2013youmove}.
\revise{In order to allow for}{To allow} use by dancers without technical skills \revise{and}{or} ownership of such specialized hardware, a support system should offer quick access to moments where their dances are not well\revise{ }{-}synchronized \revise{and should}{while} utiliz\revise{e}{ing} commodity video recording devices.
Recent work~\cite{Lee2020pose} has demonstrated a mobile system to offer feedback on dance\revise{}{s} performed by a single person.
However, the system only considers the differences in poses, and lacks the consideration of the temporal alignments of dancers' movements, which is another critical component that creates visual aesthetics seen in synchronized dancing.

We present \systemname{}, which provides support for synchronized dance practices \revise{through}{using} computer vision technology and interactive visualization.
The main advantage of \systemname{} is to offer \revise{a }{}quick access to the portions of recordings where the system considers that dances are not well\revise{}{-}synchronized in an in-situ manner.
% \zhongyi{Three practice one }{}
Unlike existing work \cite{Lee2020pose, Anderson2013youmove, chan2010virtual}, \systemname{} considers both pose similarity among dancers and the temporal alignment of motions, another key element \revise{for}{of} synchronized dancing for inferring the degree of synchronization.
\systemname{} then visualizes them as 1D heatmaps (Figure \ref{fig:overview} right).
In this manner, dancers \revise{can}{are presented with an} overview \revise{}{of }the performance evaluation estimated by the system.
They can quickly navigate to moments where the degree of synchronization is low, and \revise{deeply}{further} investigate how \revise{they can}{to} fix \revise{}{the }issues.
We envision that \systemname{} can mitigate the existing isolation \revise{between}{of} practices \revise{and}{from} review\revise{}{s} \revise{with}{using} recordings.

This work offers the following contributions:
\begin{itemize}
    \setlength{\itemsep}{0cm}
    \item A formative study that summarizes existing practice procedures and features desired in support systems for synchronized dancing\revise{,}{;}
    \item Development of two methods to quantify the degree of synchronization from synchronized dance videos captured with a normal RGB camera in terms of both pose similarity and the temporal alignment of dancers' movements\revise{,}{;}
    \item Design of an interactive system for synchronized dancing practice support\revise{,}{;} 
    \item System evaluations on the accuracy and the robustness of our two quantification methods\revise{,}{;} and
    \item A qualitative user evaluation to confirm the benefits of SyncUp and uncover its potential use\revise{}{s} in actual practices. 
\end{itemize}

In this paper, we present \revise{our explorations to}{an exploration of how we} design\revise{}{ed}, buil\revise{d}{t} and evaluate\revise{}{d} SyncUp.
We first conducted formative studies to understand current practices and problems in synchronized dance practices and derive desired features for \systemname{}.
We developed vision-based approaches to identify pose similarity and temporal alignment of motions using videos captured with a normal RGB camera.
This paper \revise{then }{}reports the results of our system and user evaluations, and discusses our findings as well as future directions for improvements. 

%% file: contents/related.tex
\section{Related Work}

\subsection{Dance Practice Systems}
Prior research \revise{has }{}uncovered the complexity of dance learning~\cite{BLASING2012Neuro, karpati2015dance}, stimulating the \revise{explorations}{development} of practice systems for dancers~\cite{Raheb2019Dance, Camurri2016System}. 
Many \revise{of }{}such systems support practice\revise{s}{} \revise{through}{with} perception enhancement~\cite{chan2010virtual, Kyan2015An}.
Drobny et al.~\cite{Drobny2009Saltate} found that a common challenge for dance learners was \revise{to synchronize}{synchronizing} their movements with the underlying rhythm of the music.
To address this issue, they created Saltate, which provides beginners with drum-sound feedback at beat onsets.
Nakamura et al.~\cite{Nakamura2005multimodal} employed a movable robot to address the depth ambiguity problem of dance learning videos that \revise{can }{}only convey 2D information to dancers.
The robot \revise{can move}{moves} forward or backward according to \revise{the }{}motion\revise{ }{-}depth information.
Their research allowed dancers to improve their understanding of their movements in a 3D space.
YouMove~\cite{Anderson2013youmove} is an augmented-reality mirror that supports dance learning by visually comparing the poses of
a student and instructor at several key dance moments.
The system also provides quantitative performance evaluations with detailed notes from the instructor to help the student understand how to improve their dancing.

Existing research has also revealed design guidelines for interactive dance practice support systems~\cite{CiolfiFelice2018Knotation, Hsueh2019Understanding, Riviere2018How}.
\revise{Through}{From} comparative studies, Trajkova et al.~\cite{Trajkova2018Takes} found that the efficacy of feedback modalities (e.g., verbal or visual) in augmented mirror systems \revise{can }{}depend\revise{}{s} on the dancer's expertise level.
Riviere et al.~\cite{Riviere2019Capturing} compared two dance decomposition methods (\revise{decomposition}{i.e.,} led by dancers or instructors) for dance learning.
They found that decomposition performed by instructors is appropriate for introductory-level students whereas\revise{}{,}
decomposition by dancers is preferable for experts.

These projects only use\revise{}{d} pose information for dance performance evaluations, and the temporal alignment of dance movements \revise{has not been}{were not} fully considered.
An interactive system that utilizes commodity devices (i.e., smartphones or tablets) and computer vision methods to support synchronized dancing practice can liberate users from specialized equipment (e.g.,a depth camera and an augmented-reality mirror), and expand use cases.

\subsection{Vision-based Human Motion Analysis}
Accurate human pose detection is an active research area in the computer vision field.
Researchers have developed real-time robust pose detection methods using depth cameras~\cite{Krull2015Learning,Shotton2011Real,  Shotton2013Efficient}.
% \koji{}{cite some work like Kinect CVPR paper done by MSR cambridge.}
Additionally, recent breakthroughs in computer vision~\cite{Alex2012ImageNet} have established robust algorithms to locate human skeletons in images, including those captured \revise{with}{by} a typical RGB camera~\cite{Cao2018Realtime, Fang2017RMPE, Xu2018MonoPerfCap, Zhang2019pose}.
The key idea behind these methods is to train convolutional neural networks (CNNs) using large-scale fine-labeled datasets~\cite{andriluka14HumanPose, Lin2014coco}.
OpenPose~\cite{Cao2018Realtime}, one of the earliest CNN-based real-time multi-person pose estimators, \revise{can }{}predict\revise{}{s} the pixel positions of 18 different human body parts (\textit{keypoints}) in a given frame.
Thus, developers can easily construct body skeletons \revise{with}{using} these predictions.
\systemname{} employs AlphaPose~\cite{Fang2017RMPE}, another well-known multi-person pose estimator, because of its superior performance \revise{compared to}{vs.} OpenPose. 

\subsection{Vision-based Pose Similarity Estimation}
Pose similarity estimation plays a critical role in various applications, including action recognition~\cite{du2015hierarchical, liu2016spatio}, motor skill learning~\cite{Anderson2013youmove, chan2010virtual}, and motion retrieval~\cite{Sedmidubsky2013KeyPose, Sun2020View}.
One basic pose similarity estimation approach uses the Euclidean distances between body parts in two poses. 
For example, Chan et al.~\cite{chan2010virtual} created a virtual reality dance practice system that employed a simple threshold-based method using the Euclidean distance of each body part between the reference dancer and learner. 
However, Chen et al.~\cite{chen2010learning} noted that such features were insufficient to evaluate more complex actions.
Therefore, they introduced a collection of high-dimensional pose features (1,683 dimensions) to describe human motions and an algorithm to learn the similarity metric based on \revise{the }{}Mahalanobis distances, which was found robust when using \revise{such }{}these high-dimensional features.
Recently, researchers~\cite{Shi2019Two, yan2018spatial} started to encode the semantics of pose similarities using an implicit model (i.e., a graphic CNN) instead of manually\revise{-}{ }crafted features.
However, these methods require a large dataset, and their direct application to synchronized dancing is not yet feasible.

Although pose similarity estimation is a critical component in dance practice support systems, \revise{much}{most} prior work \revise{still }{}relies on the summation of the Euclidean distances of body parts.
These methods implicitly assume uniform\revise{al}{} weights on body parts, and their results may not match \revise{}{well }with human perception\revise{ well}{}.
For instance, the angles of arms may contribute to the perceived pose similarity of dancers more strongly than those of legs.
Thus, further explorations are necessary to build a pose similarity quantification method for synchronized dancing.

\subsection{Visual Beats}
Existing computer vision research has explored methods for intelligent systems to understand the tempo of human motions.
One application of these methods is to synthesize dance-like motions~\cite{Xie2020Dance, li2021learn, zhuang2020music2dance}.
Davis and Agrawala~\cite{Davis2018VisualBeat} developed the concept of a visual beat, which involves the rhythmic patterns of \revise{movements of objects}{object movements} in a video. 
Their system \revise{can synthesize}{synthesizes} dance-like motions by matching visual beats with the beats of the background music.
Lee et al.~\cite{Lee2019Dancing} developed a system that generates various types of artificial human-dancing videos based on the background music style, such as ``Ballet" or ``Zumba".
Their method uses a generative adversarial network generator~\cite{Goodfellow2014Generative} that includes a feature \revise{that }{}represent\revise{s}{ing} the music genre to recurrently create a sequence of video frames.

Our work applies visual beats to synchronized dance performance evaluations.
Instead of synthesizing dance-like motions, the primary objective of this work lies in quantifying the temporal alignment of dancer motions. 
In particular, \systemname{} extracts the visual beat features from dancers' movements\revise{,}{} and computes how their timings are aligned (i.e., whether a movement is ahead of or behind the reference motion).

%% file: contents/formative.tex
% Table generated by Excel2LaTeX from sheet 'Sheet1'
\begin{table*}[t]
  \centering
  \small
  \caption{Online survey results.
  Features are sorted by their average scores from highest to lowest. Six features (F1, F3, F4, F5, F7 and F8) were implemented in the current \systemname{} system. Note that we excluded F2 and F6 in this work because they are already well supported in existing systems.
  }
    \begin{tabular}{p{13em}p{30em}c}
    \Hline
      Feature &
      Description &
      Mean (SD)
      \\ \hline \hline
      \textbf{F1. Pose correction feedback} &
      Provide detailed textual and/or visual instructions on how to fix dancer out-of-sync motions and poses. &
      4.66 (0.55)
      \\ \hline
      F2. Slow practice mode &
      Allow dancers to practice a portion of or the entire dance at a slower tempo. &
      4.44 (1.06)
      \\ \hline
      \textbf{F3. Mistake summary} &
      Automatically summarize the moments where dancers made mistakes. &
      4.34 (0.81)
      \\ \hline
     \textbf{F4. Temporal alignment feedback}&
     Visualize the synchronization degree between dance and music. &
     4.31 (0.81)
     \\ \hline
      \textbf{F5. Leader mode} &
      Allow a team leader to record the reference video, and other dancers \revise{can}{to} compare it to their own dancing. Dancers can also visually compare their dancing and ground truth side by side using two video streams. &
      4.21 (1.01)
      \\ \hline
      F6. Focus mode &
      Allow dancers to repeatedly practice a specific portion of the dance. &
      4.17 (1.14)
      \\ \hline
      \textbf{F7. Comparison against previous practices} &
      Allow dancers to compare their current dancing to past data. &
      4.10 (1.08)
      \\ \hline
      \textbf{F8. Detection of pose differences} &
      Detect and highlight differences in poses among multiple dancers. &
      3.97 (0.87)
      \\ \hline \hline
      Hand pose tracking &
      Track dancer hand poses during dancing and provide feedback. &
      3.65 (1.20)
      \\ \hline
      Dance scoring &
      Scores represent the degree of pose similarity, motion similarity and facial expressions. &
      3.55 (1.12)
      \\ \hline
      Detection of differences in motions between poses &
      Detect and highlight differences in transition between two consecutive poses among multiple dancers. &
      3.55 (1.12)
      \\ \hline
      Facial expression detection &
      Detect dancer facial expressions and provide feedback (e.g., ``smile more"\revise{,}{} or ``show your face"). &
      3.38 (1.12)
      \\ \hline
      Detection of motion stability &
      Detect dancer stability\revise{,}{} and provide feedback. &
      3.34 (1.01)
      \\ \Hline
    \end{tabular}%
  \label{survey}%
\end{table*}%

\section{Formative Studies}
% What kind of things we should design.
% Taken from the prior work (*) or interview.
% Take their justification. 

% What groups need for successful dance.
% Check other dance support systems paper. Check how they justify the user interface design.
% we have a cooperation 

To inform our interface design, we conducted focus groups to understand how dancers practice synchronized dancing.
Additionally, we conducted an online survey to prioritize features to be implemented in SyncUp.

\subsection{Focus Groups}
We first held focus groups to understand the current practices of synchronized dancers and derive potential system features.
We recruited three dance groups \revise{in}{at} our university to participate in the focus groups.
Each dancer group consisted of 4\revise{ -- }{--}10 people who regularly perform\revise{}{} synchronized dancing at university events (e.g., school festivals) and external occasions (e.g., dance competitions).
We began our focus group by asking ice-breaker questions (e.g., what type of dancing they perform\revise{ed}{} and how often they practice\revise{d}{}), and then we asked our participants to share their practice approaches and current challenges.

We extracted the common practices and problems observed in our focus groups\revise{,}{} and derived the following major findings: 

\begin{itemize}
    \setlength{\itemsep}{0cm}
    \item All groups used video recording (mostly \revise{using }{}smartphones) for their practices. These videos were useful for finding small but important mistakes in their dancing. In addition, the participants commented that recorded videos allow them to review their dancing from the audience perspective.
    \item All groups preferred checking their recording only after finishing their entire practices (e.g., after going back home).
    % \revise{}{ although \xxx dance groups in our focus group explicitly mentioned that they desired to do so during their practice}
    One group had never checked their video\revise{}{s} when they were together because it was time consuming.
    \item In all groups, dancers first practiced individually\revise{, and then}{; then,} they practiced as a group. In this manner, they \revise{can}{could} focus on improving their synchronization instead of learning movements and poses during group practices.
    \item All groups typically selected a leader \revise{who is }{}responsible for providing feedback to the other members (\revise{}{i.e., }followers) in their group practices. However, they noted that the leader \revise{cannot}{could not} always identify all mistakes. 
    % \item \koji{}{add a sentence here about when they replayed their captured videos.}
    % \zhongyi{\item 
    % One group always replayed their practice videos after finishing the practice, using a standard video player interface. They also explained that this process was time-consuming.
    % % about the inconvenience of a standard video player to locate their mistakes moments.
    % % For example, they needed to explore in the progress bar, looking for target 
    % % , such as looking for some problematic movements by repeated manipulate the progress bar.
    % }{New}\koji{}{what about the other groups? When did they check their videos?}
    \item One group manually created a sheet that summarizes which parts of their dances require\revise{}{d} additional practice after repeatedly reviewing their videos.
\end{itemize}

% \begin{itemize}
%     \item Derive a set of features from prior work and existing dance practice systems that our target users can potentially desire.
%     \item narrow down the set into 13 features after discussions among authors including a person who have extensive experience in group dancing. Summarize the 13 features as a table.
%     \item run a survey, how long, how many you got, who responded.
%     \item results as a table starting from most desired features
%     \item brief discussions on what to implement and what not to. Justifications on why we did not implement some features.
% \end{itemize}

We found that video recording is a common approach for synchronized dancing practice.
However, focus group participants \revise{relied}{tend to rely} on the manual inspection of these recorded videos to identify dance segments requiring further practice.
These results \revise{were}{are} in line with our motivation to design an interactive system that supports synchronized dance practices \revise{with}{using} commodity devices.
% A summary sheet covering mistakes observed in practices would offer a quick access to the moments where dancers would need additional practices.
% \zhongyi{does the following sentence implies creating the sheet is the core of our system? 
% Basically, we can create an excel-like sheet, but currently the system only outputs an website interface that summarize an evaluation of the dance.}
% However, creation of such summary sheets currently relies on manual work, and would not scale.

\subsection{Online Survey}
We \revise{then}{later} conducted an online survey to understand which functionalities are most desired by synchronized dancers.
We generated 13 potential features (Table~\ref{survey}) based on the results of our focus groups to address participant concerns regarding the time-consuming aspects of their practices.
We conducted an online survey to examine the importance of each feature from the user perspective.
The survey asked participants to rate each feature on a \revise{5-}{5-Point} Likert scale concerning the potential usefulness to their practice (5: they would definitely use the feature\revise{ -- }{; }1: they would never use the feature).
We recruited 29 individuals (\revise{6}{six} males and 23 females) with different dancing engagement levels\revise{,}{} from casual dancers to professionals.

% Our focus groups included discussion about time-consuming aspects of their practices, requirements of future supporting systems, and features they think would be useful.
% We then summarize their ideas into 13 possible interface feature proposals as shown in Table~\ref{survey}.
%Carla: I rewrote this paragraph as follows, so it focuses on this study and not the previous one.

% All candidates of our survey are group dancers from various levels: 1) beginners: who have only practiced once or twice for several performance; 2) amateurs: who frequently attend some dancing activities; 3) professions: who regard group dancing as their jobs. 
%There are 29 candidates of our surveys: 6 males and 23 females.
%We provides them with 13 feature proposals with 5 Star rating principle: 5 indicates that the feature is useful and s/he will use it most of the time.

\subsection{Survey Results and Discussion}
Table~\ref{survey} summarizes our results, including the mean rating and standard deviation of each feature.
All \revise{the }{}features were considered useful; the lowest average score was greater than 3.0.
However, implementing all \revise{the }{}proposed features \revise{may}{would} overwhelm our target users and hinder the core value of our system.
In addition, certain features \revise{can}{may} be error-prone due to the limitations of existing computer vision technology (e.g., hand pose tracking).
Furthermore, we decided to exclude features that are already well supported by existing systems (e.g., F2 and F6) and focus\revise{}{ed} on designing quanti\revise{fied}{tative} feedback on dance performances.
We thus narrowed down the features to those that are critical and sufficiently robust with \revise{the }{}existing technolog\revise{y}{ies} for our prototype.
We excluded features with scores of 3.65 or below as there was a substantial gap (0.32) there.
This led us to focus on the six features in Table~\ref{survey} (\revise{}{i.e., }F1, F3, F4, F5, F7, and F8).
%: pose correction feedback, slow practice mode, mistake summary, tempo synchronization feedback, teacher/leader mode, section loop mode, comparison against previous dancing, and detection of pose differences.

% Among all the proposals, we decide to focus on those with averages score higher than, or equal to, 4 as our interface design targets.
% Such high scores show higher requirements from a large majority of our potential users.
% For example, our preliminary study finds that a certain number of amateur-level groups have some person responsible for manually creating a "mistake summary" sheet.
% This person has to check their dancing recording videos multiple times and take screen-shots of a person's wrong pose as well as others' poses as a comparison.
% A series of such mistake with other comments compose one mistake summary for one group member.
% An automatic generation of such summary can dramatically support their work efficiency.

%This survey constitutes the 7 features our system would have, excluding the "slow mode", the feature that various prior systems [add reference] have already contain.

% Besides, though ``Slow Mode" is ranking high, previous researches have already made various systems which contain this function.
% Therefore, 

%   \caption{The result of our interface feature survey.
%   The features are sorted by their average scores from high to low values.
%   }

%% file: contents/interface.tex
\section{SyncUp Interface}
\systemname{} \revise{was}{is} designed as a \revise{W}{w}eb-based interface that can run on laptops, smartphones, and tablets.
All \revise{the }{}features except the spotlight view (explained in Section~\ref{spotlightview}) are contained in one \revise{W}{w}eb page (F3).

We envision that dancers \revise{would}{will} capture their multiple \revise{sessions of practices}{practice sessions} consecutively at one time, and SyncUp \revise{would}{will} then offer feedback when they \revise{were taking}{take} \revise{a }{}short break\revise{}{s} between practices.
When each practice session \revise{is finished}{finishes}, the system \revise{would}{will} upload the video for analysis while a mobile device \revise{would be capturing}{captures} the next session.
In our current implementation, SyncUp requires approximately 85 \revise{seconds}{s} to analyze a \revise{one-minute-long}{1-min} video with three dancers \revise{with}{at} 30 \revise{frames per second}{fps} and \revise{in the resolution of}{at} 360p \revise{}{resolution}.
In addition, providing feedback in a post-hoc manner (i.e., after finishing a practice session rather than real time \revise{in dancing}{during the practice}) is in line with the findings by Trajkova et al.~\cite{Trajkova2018Takes}.
SyncUp \revise{would }{}highlight\revise{}{s} segments where synchronization is deemed low by the system, and dancers can discuss whether they \revise{would like}{want} to \revise{}{add }practice \revise{more on the}{in} corresponding portions of their dance program.
In this manner, SyncUp \revise{can }{}alleviate\revise{}{s} the isolation of practices \revise{and}{from} review\revise{}{s} using video recordings\revise{that we}{, as} observed in our formative studies.

\begin{figure}[t]
    \centering
    \begin{subfigure}[b]{0.45\linewidth} 
        \includegraphics[width=0.99\linewidth]{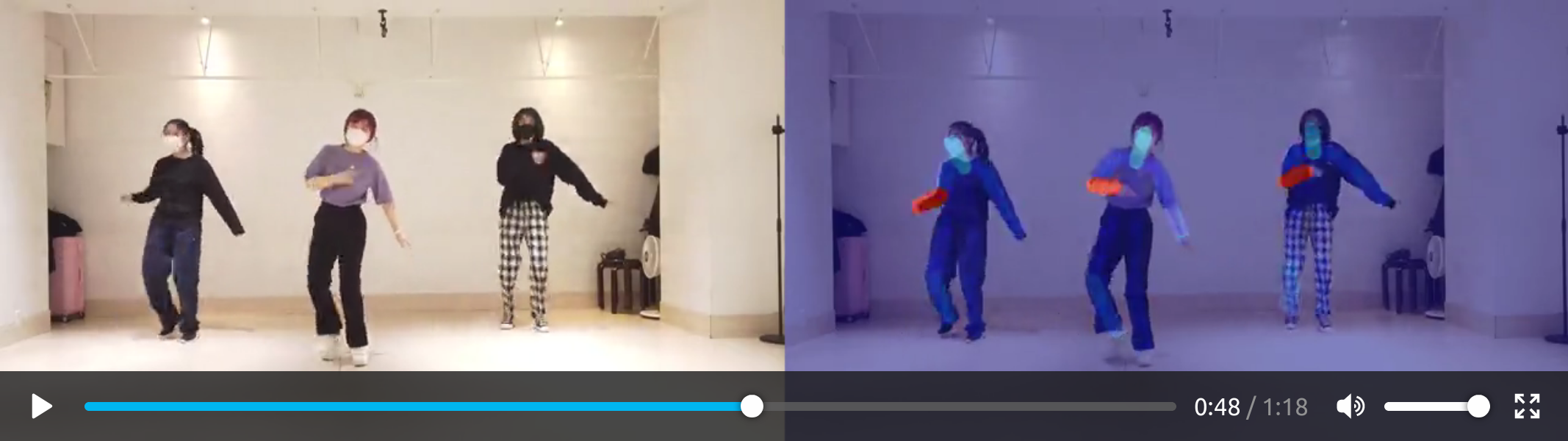}
        \caption{Side-by-side video replay with heatmap overlays in the group practice mode.}
        \label{fig: interface_video_group}
    \end{subfigure}
        \centering
    \begin{subfigure}[b]{0.45\linewidth} 
        \includegraphics[width=0.99\linewidth]{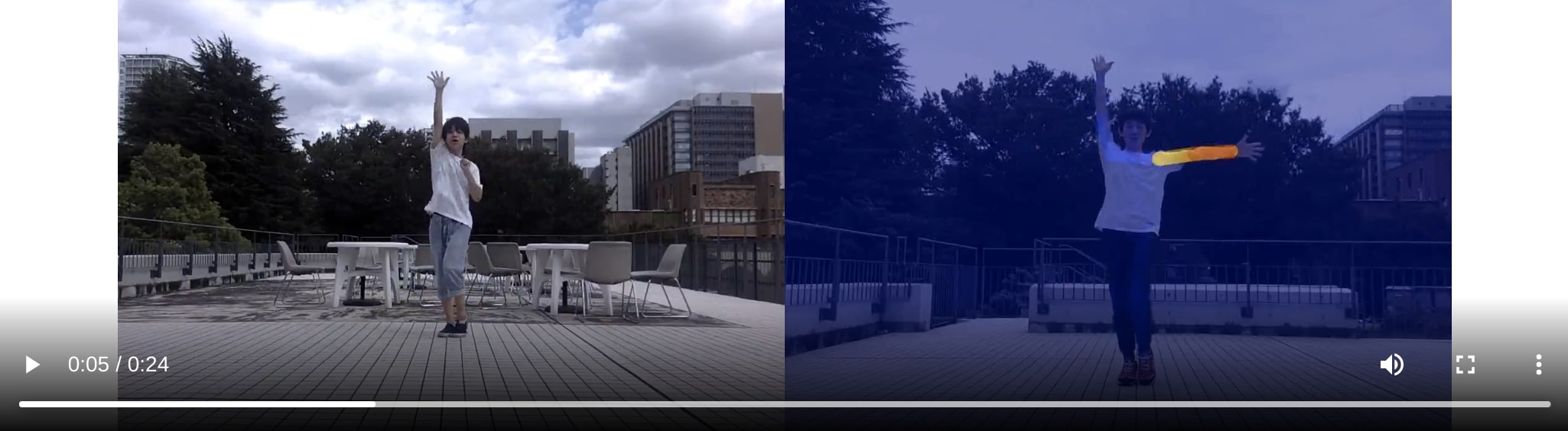}
        \caption{Side-by-side video replay with heatmap overlays in the individual practice mode.}
        \label{fig: interface_video_indv}
    \end{subfigure}
    % \\

    \begin{subfigure}[t]{0.8\linewidth}
        \centering
        \includegraphics[width=0.99\linewidth]{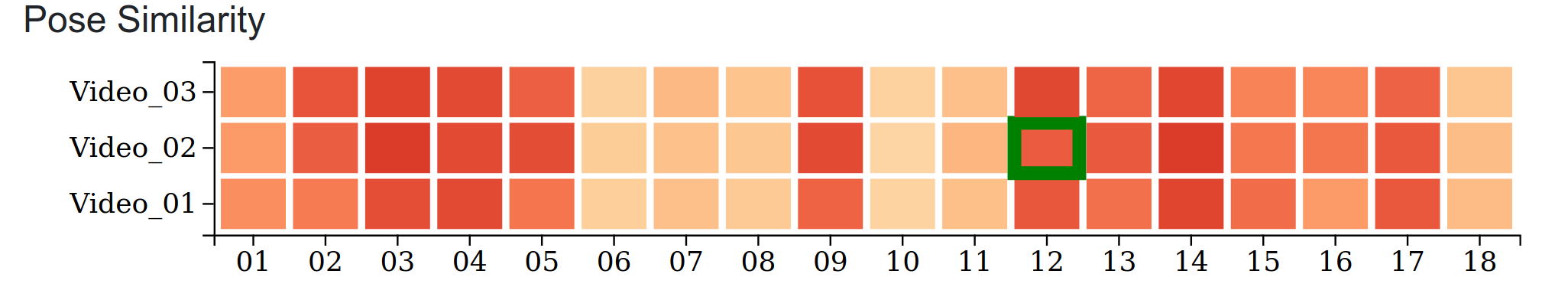}
        \caption{Pose similarity visualization. 
        Dance segments with low pose similarity are indicated by a dark red color.
        }
        \label{fig: interface_pose}
    \end{subfigure}
    \begin{subfigure}[t]{0.8\linewidth}
        \centering
        \includegraphics[width=0.99\linewidth]{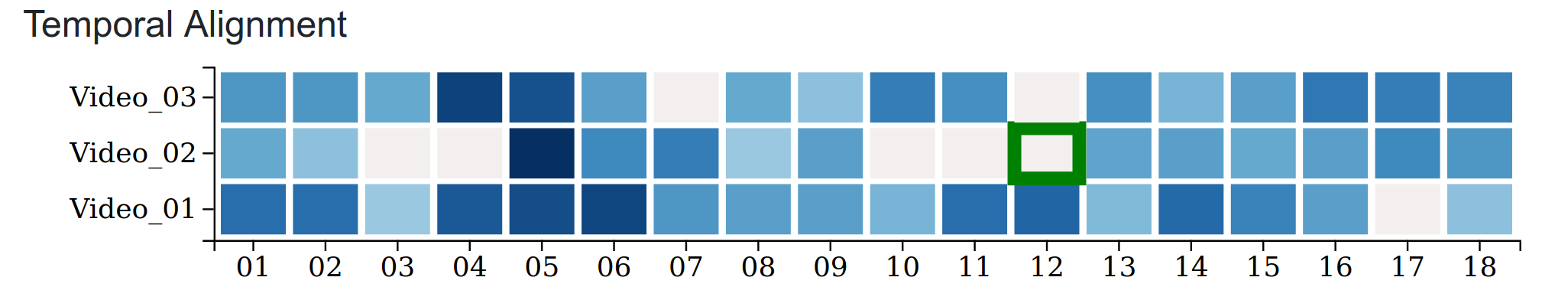}
        \caption{Temporal alignment visualization. 
        Dance segments where dancers' movements are not aligned are represented by a dark blue color.
        }
        \label{fig: interface_temp}
    \end{subfigure}
    \caption{Interactive visualizations in \systemname{}. 
    \systemname{} presents a side-by-side video (either \ref{fig: interface_video_group} or \ref{fig: interface_video_indv}) at the top of the interface. 
    The interface also includes 1D heatmaps of multiple practice videos. 
    The values in the horizontal axis in both \ref{fig: interface_pose} and \ref{fig: interface_temp} show\revise{s}{} the indices of video segments (8 beats long).
    Green rectangles highlight the corresponding segment played in the side-by-side video replay view.
    Both \ref{fig: interface_pose} and \ref{fig: interface_temp} show the analysis results for \ref{fig: interface_video_group}.
    }
    \label{fig: Interface} 
\end{figure}

\subsection{Side-by-Side Video Replay}
The main feature in \systemname{} is the side-by-side video replay with heatmap overlays.
The contents in the video replay view \revise{are different}{differ} in the two practice modes \revise{}{offered by }SyncUp\revise{ offers}{}: \revise{the }{}group \revise{practice mode }{}and individual practice mode\revise{}{s}.

\subsubsection{Group Practice Mode}
The group practice mode is utilized when dancers are physically together\revise{ and are}{} practicing a synchronized dance. 
In the group practice mode, \systemname{} offers a side-by-side view of the original practice video and the \revise{same video}{one} with heatmap overlays that visualize the degree of similarity of dancer poses at a particular time (F1).
In the example of Figure~\ref{fig: interface_video_group}, the heatmap overlays highlight dancers' arms in red, suggesting that this body part is not well synchronized.
This visualization helps dancers quickly identify which body parts require adjustment in future practices.

\subsubsection{Individual Practice Mode}
The individual practice mode is used when dancers wish to practice separately.
Our focus groups suggested that this is common \revise{in}{during} the early stages of practices.
In this practice mode, \revise{there is typically a}{one} dancer \revise{who }{}is \revise{}{typically} responsible for teaching the other\revise{ dancers}{s}.
In \revise{the following content of the}{this} paper, we refer to this dancer as the ``leader'' and the other\revise{ dancers}{s} as \revise{the }{}``followers''.
The side-by-side video replay shows the video of the leader on the left and \revise{the video}{that} of the follower with heatmap overlays on the right (Figure \ref{fig: interface_video_indv}).
\revise{The alignment}{Alignment} between the two videos is performed through the background music alignment~\cite{ellis2014fingerprint}.
This mode \revise{also }{}allows the leader to record \revise{her}{their} dances as the reference data to support followers' practices (F5).

\subsection{Pose Similarity Visualization}
Pose similarity quantifies how well the poses of multiple dancers are synchronized in a particular video frame. 
In addition to the heatmap overlays, \systemname{} also offers a 1D heatmap to visualize the average score of pose similarity (F8) in each segment of the dance video\revise{,}{} derived by our algorithm (explained later).
A darker red color indicates a lower degree of synchronization.
We define the length of a segment as 8 beats long because it is commonly used in dance practices.
Multiple rows of \revise{the }{}1D heatmap\revise{}{s} present a comparison of the synchronization scores among multiple practice videos.
This helps dancer\revise{}{s} review how their pose similarity \revise{has been improved}{improves} over multiple rounds of practice\revise{s}{} (F7, Figure~\ref{fig:overview} right).

When the user clicks a block o\revise{f}{n} the heatmap, it activates a replay of the corresponding segment in the side-by-side video replay\revise{,}{} at a play speed specified by the user.
In addition, when \revise{the}{a} user \revise{is replying}{replays} the video, the green highlight in the heatmaps is also updated to inform the location of the corresponding segment.

\subsection{Temporal Alignment Visualization}
\label{sect: interface_tempo}
Pose similarity considers the degree of synchronization at a video frame level.
Another aspect that contributes to \revise{the }{}synchronization is the alignment of the movement timing.
For example, if one dancer \revise{is raising her}{raises his/her} right hand \revise{up }{}at the same time that another dancer \revise{is raising her}{raises his/her} left leg\revise{ up}{}, the poses of two dancers \revise{are different}{differ}, but the timing of their motions is synchronized.
We thus need to \revise{}{also }consider whether dancers' motions are aligned in the temporal domain\revise{ as well}{}.

It is challenging to directly detect how the movements of the dancers are aligned with the music without human annotations.
Such annotations \revise{would be}{are} very tedious \revise{for dancers}{}.
Our system therefore determines how the timing of the followers' movements (\revise{}{i.e., the }two dancers on the side \revise{in}{of} Figure~\ref{fig: interface_video_group}) is synchronized with that of the leader (\revise{}{i.e., }the central dancer in Figure~\ref{fig: interface_video_group}).
Dancers \revise{can name}{identify} the leader by choosing \revise{her}{a} skeleton from \revise{a}{the} menu.
\systemname{} then visualizes the degree of the temporal alignment of their movements in a 1D heatmap, as shown in Figure~\ref{fig: interface_temp} (F4).
A darker blue color suggests a larger difference in terms of the movement timing.
Similar to the pose similarity visualization, users can click a block of the heatmap to navigate to the corresponding segment.

\subsection{Spotlight View}
\label{spotlightview}

In addition to the default view shown in Figure~\ref{fig: Interface}, \systemname{} \revise{has}{offers} another interface\revise{ called}{:} the spotlight view.
\revise{The spotlight}{This} view offers a collection of segments that are sorted by the synchronization score calculated from the pose similarity and temporal alignment in \revise{the }{}ascending order.
The segment \revise{at the top of}{atop} this view is the portion of the video in which the system considers that a dance \revise{was the most ill-synchronized}{had the worst synchronization}.
This view is intended to offer direct access to moments where additional practices \revise{would}{may} be necessary by liberating users from inspecting with the full-length footage.

%% file: contents/method.tex
\begin{figure*}
  \begin{subfigure}[b]{0.99\textwidth}
    \includegraphics[width=\textwidth]{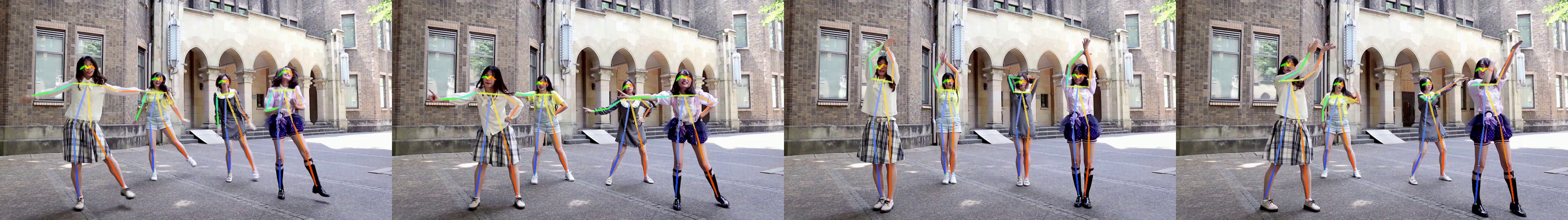}
    \caption{Skeletons constructed based on pose detection with AlphaPose~\cite{Fang2017RMPE, xiu2018poseflow}. The four frames are randomly selected.}
    \label{pose}
  \end{subfigure}
  \begin{subfigure}[b]{0.99\textwidth}
    \includegraphics[width=\textwidth]{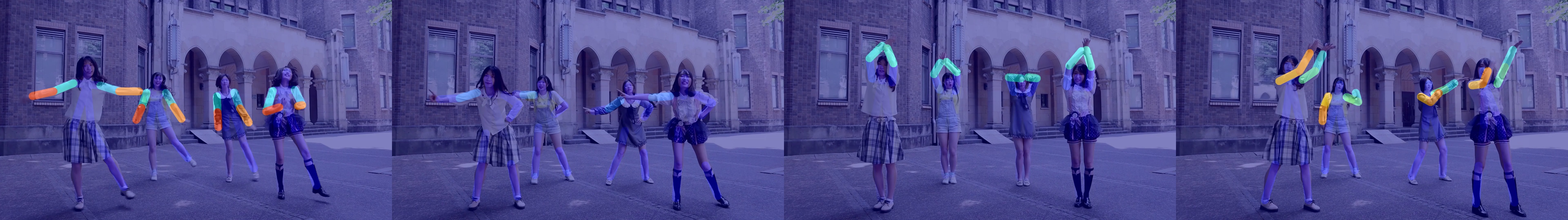}
    \caption{Body-part-level Pose Distance (BPD) visualization with heatmap overlays.}
    \label{vis}
  \end{subfigure}
  \caption{
%   \zhongyi{Change the figure, misaligned, changing}
  Example results of our pose similarity quantification method.
  }
  \label{result}
\end{figure*}

\section{Dance Performance Quantification}

\systemname{} executes two kinds of analyses to quantify the degree of synchronization in dances: pose similarity and temporal alignment.
Pose similarity analysis examines how the poses of multiple dancers are synchronized with each other in a given frame.
Temporal alignment analysis quantifies the timing differences of motions among multiple dancers, indicating how well synchronized a dancer's motion is with others at a specific time interval.
Our interface ultimately visualizes the results of these analyses, as shown in Figure~\ref{fig: interface_pose}.

\subsection{Video Pre-processing}
\subsubsection{Video Segmentation by Music Beats}
Counting eight beats in a loop is a common dance practice technique.
Referring to this technique, \systemname{} divides a practice video into a series of segments, each of which is 8 beats long.
We use Ellis's method~\cite{ellis2007beat} to estimate the tempo of the music and segment the video. 

\subsubsection{Dancer Pose Detection}
We use AlphaPose~\cite{Fang2017RMPE}, a multi-person pose estimator, for identifying dancers' poses.
In AlpahPose, a person's pose is represented by a list of pixel positions \revise{of}{having} 18 keypoints ($p$)~\cite{Cao2018Realtime, Lin2014coco}\revise{}{,} similar to OpenPose~\cite{Cao2018Realtime}.
These keypoints are the 2D pixel locations of \revise{the two}{a person's} eyes, \revise{two }{}ears, \revise{a }{}nose, \revise{a }{}neck, \revise{two }{}shoulders, \revise{two }{}elbows, \revise{two }{}wrists, \revise{two }{}hips, \revise{two }{}knees, and \revise{two }{}ankles\revise{ of a person}.
Figure~\ref{pose} displays the overlays of skeletons created from the AlphaPose keypoints on example recorded dancing videos.
We did not formally experiment with the accuracy of pose detection \revise{as}{because} it is out of the scope of this work.
However, we found that AlphaPose successfully detected dancers' poses except in the case where a person was largely occluded. 
\subsubsection{Dancer Tracking}
After \revise{the }{}pose detection, the system performs Xiu et al.'s method~\cite{xiu2018poseflow}\revise{ to}{, which} track\revise{}{s} each dancer between frames.
We hypothesize that the poses of the same dancer at two consecutive frames \revise{would}{will} be very similar.
\revise{This means that}{Hence,} the body part locations \revise{would}{will} also be similar between the \revise{two }{}frames.
The system first calculates the distances of all skeletons between \revise{the two }{}frames.
We define the skeleton distance between \revise{the two }{}frames as the total distance\revise{}{s} of \revise{each of}{the} 18 keypoints.
The system then finds the combination that minimizes the overall distances, resulting in the mapping of the dancers between \revise{the two }{}frames.

In rare cases, AlphaPose fails to identify the same number of dancers as in the previous frame (e.g., \revise{3}{three} dancers in the previous frame, but \revise{2}{two} in the current frame).
The system still performs the same procedure\revise{ above}{}, and it carries over the skeleton information of the unmatched dancer\revise{s}{(s)} to the current frame (i.e., the system regards that these dancers did not move at all between the two frames).

\subsection{Pose Similarity Analysis}

The algorithm first computes the degree of pose similarity 
among dancers at the body part level (\textit{Body-part-level Pose Distance}, or BPD in short), which the system uses to display the heatmap overlay (Figure \ref{fig: interface_video_group} $\&$ \ref{fig: interface_video_indv}).
The system also computes a metric called \textit{Overall Pose Similarity} (OPS), which  describes the overall similarity of poses at a given frame with the BPD values of all dancers.
SyncUp then averages the OPS values across \revise{the all the}{all} frames in each segment\revise{,}{} and presents the result \revise{in}{as} 1D heatmaps (Figure~\ref{fig: interface_pose}).
% that is used in our line plots.

\subsubsection{Body-part-level Pose Distance}
\systemname{} uses 14 keypoints (all except eyes and ears) to compute 13 body\revise{ }{-}part feature vectors.
We exclude\revise{d}{} the four features of eyes and ears because we wanted to focus on \revise{the }{}body poses rather than \revise{the directions of faces}{face directions}. 
We normalize the\revise{se}{} feature vectors into their corresponding unit vectors to preserve only the directional information of each body part.
This avoids a scale-difference problem caused by different \revise{}{human }heights\revise{ of people}{} and \revise{different distances from a camera}{camera distances}.
We use these 13 unit vectors as \revise{a}{the} source input \revise{of}{to} our algorithm to calculate the BPD value ($BPD(i,t)$) for each body part $i$ at a given frame $t$.

The system calculates the accumulated absolute difference of each body part across all dancers ($d(i,t)$) according to the formula:
\begin{equation}
    d(i,t) =\sum^{J}_{j}|\vec{v}_{j}(i,t)-\vec{v}_{R}(i,t)|
    \label{equ: bp_equ}
\end{equation}
where $\vec{v}_{j}(i,t)$ is the $i$-th unit vector ($i \in \{1,2 ... ,13\}$)  of the $j$-th person ($j \in \{1,2 ... ,J\}$), and $\vec{v}_R(i, t)$ is the $i$-th unit reference vector, equal to 
the average of the unit vectors of all \revise{the }{}dancers (i.e., $\vec{v}_R(i,t) = \frac{1}{J}\sum_j \vec{v}_{j}(i,t)$).

Note that $d(i,t)$ is an unbound variable\revise{, and it}{ that} is strongly influenced by the number of people in the analysis.
We thus use the following formula for normalization to derive $BPD$:
\begin{equation}
    BPD(i,t) =\left(\frac{d(i,t)}{J}\right)^\lambda=\left(\frac{\sum_{J}|\vec{v}_{j}(i,t)-\vec{v}_{R}(i,t)|}{J}\right)^{\lambda}
\label{equ: partialsync}
\end{equation}

A higher value of $BPD(i,t)$ means a larger discrepancy in the poses of the $i$-th body part across dancers, implying low synchronization performance.
In our heatmap overlays, we linearly convert the value of $BPD(i,t)$ to a color in the spectrum implemented in the COLORMAP\_JET class\footnote{\url{https://docs.opencv.org/2.4/modules/contrib/doc/facerec/colormaps.html}}.

\begin{figure}[t]
    \centering
    \begin{subfigure}[b]{0.3\linewidth} 
        \centering
        \includegraphics[width=\textwidth]{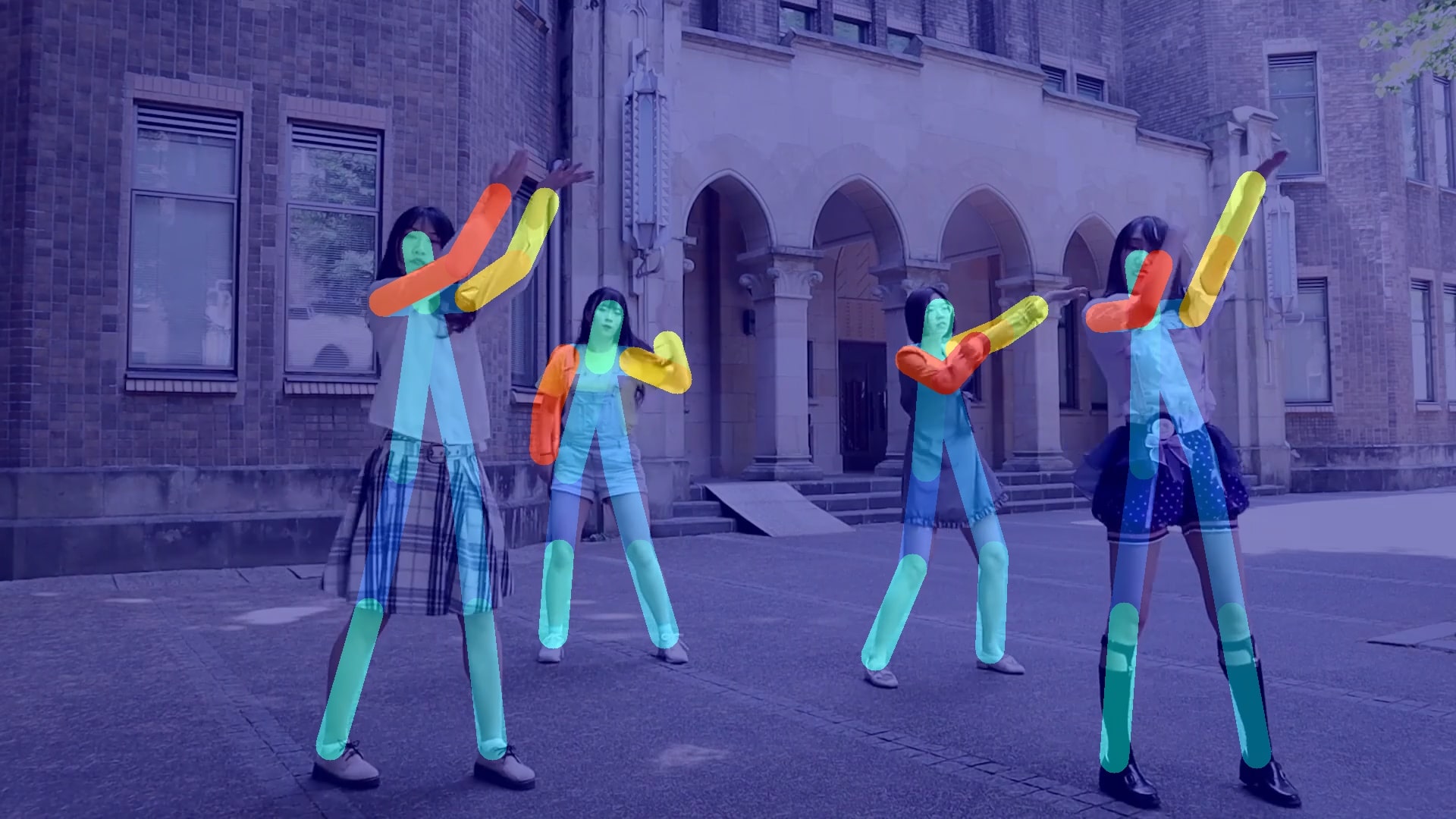}
        \caption{$\lambda = 0.5$}
    \end{subfigure}
    % \\
    \begin{subfigure}[b]{0.3\linewidth}
        \centering
        \includegraphics[width=\textwidth]{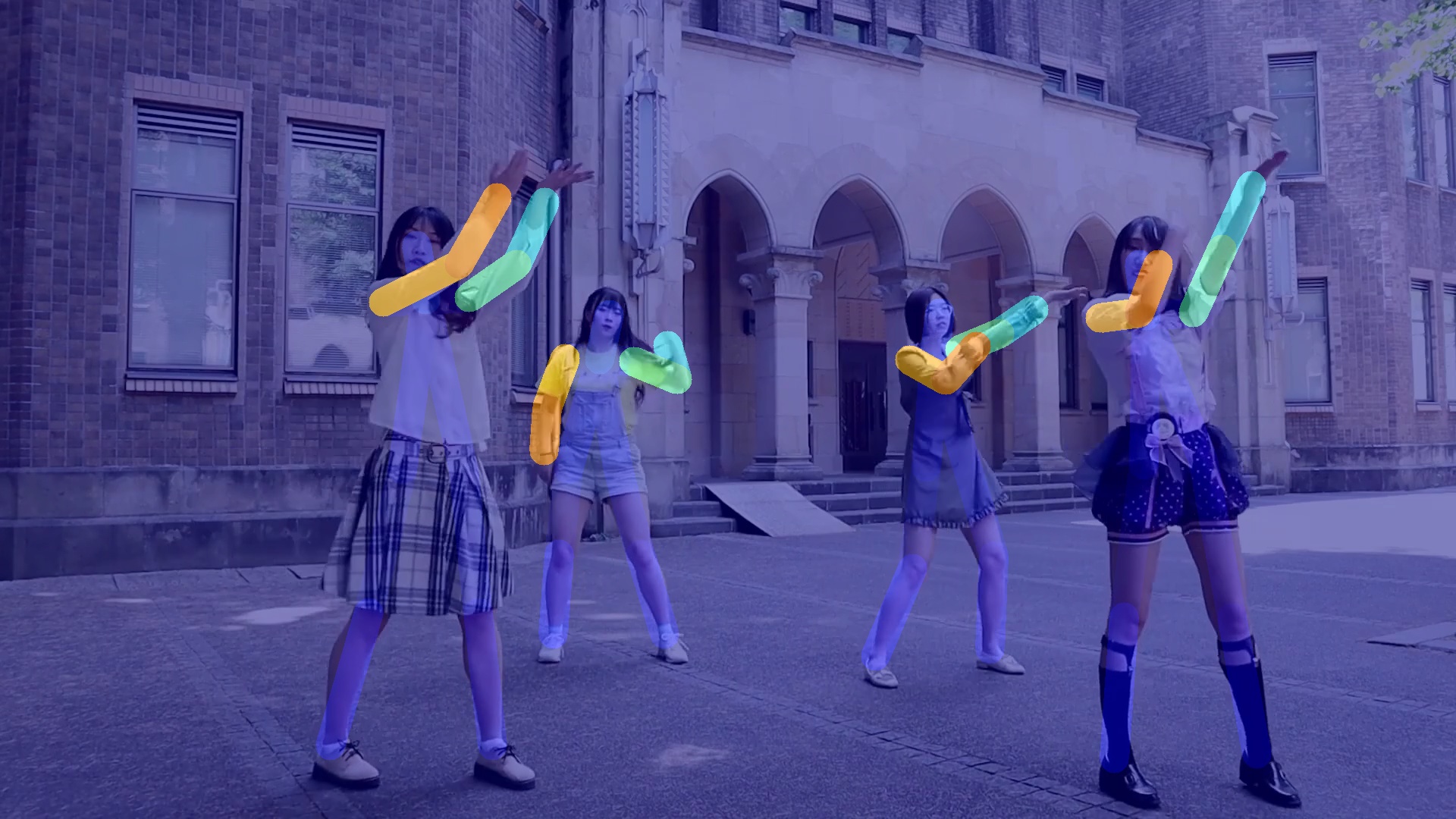}
        \caption{$\lambda = 1$}
    \end{subfigure}
        \begin{subfigure}[b]{0.3\linewidth}
        \centering
        \includegraphics[width=\textwidth]{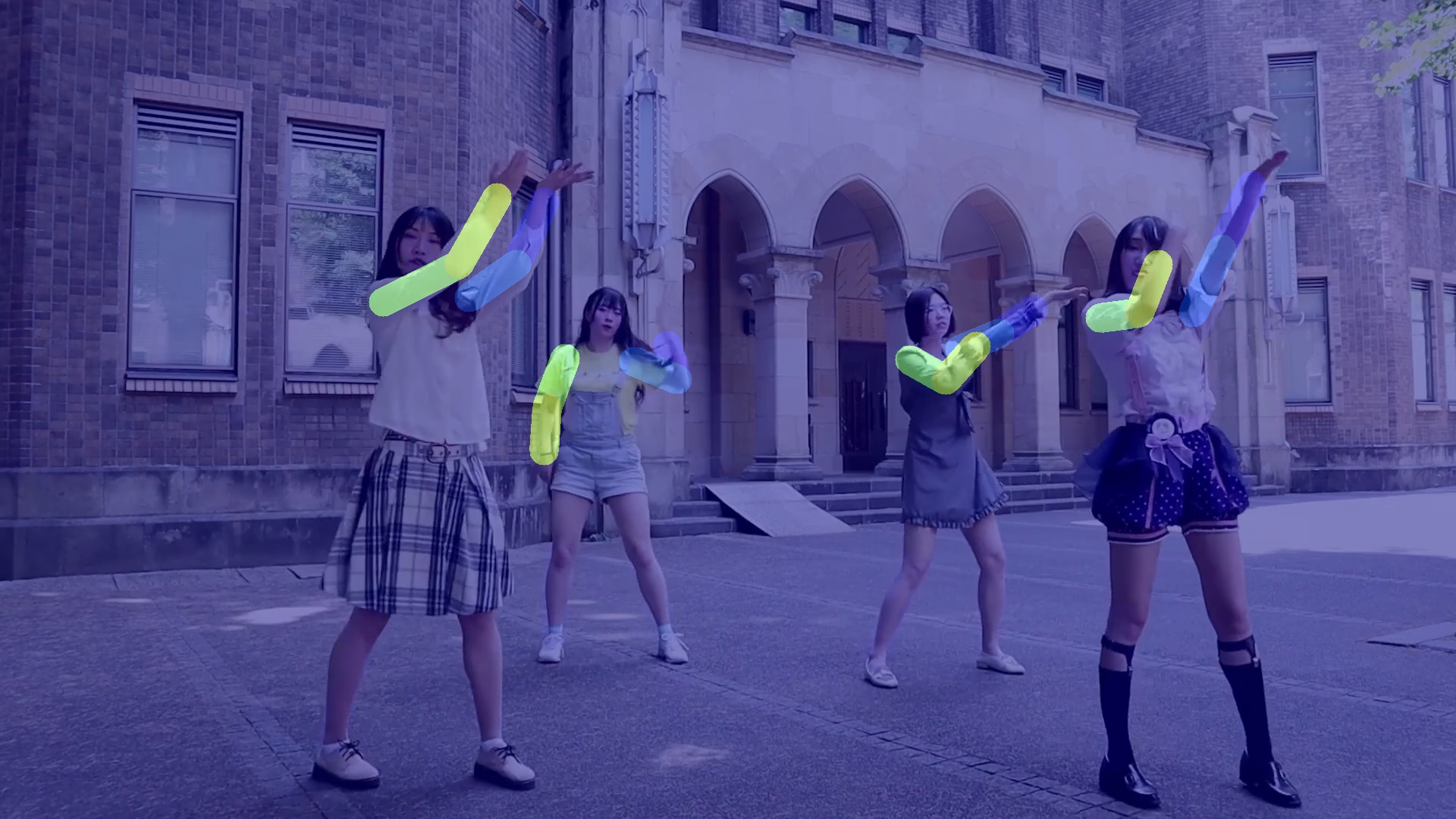}
        \caption{$\lambda = 2$}
    \end{subfigure}
    \caption{Heatmap overlay results with different $\lambda$ values.
    }
    \label{fig: lamb} 
\end{figure}
$\lambda$ is a trade-off parameter \revise{}{used }to control the sensitivity of the heatmap overlays to the discrepancy of poses.
Figure~\ref{fig: lamb} shows how this parameter can affect \revise{\systemname{}'s heatmap overlays}{the heatmap overlays of \systemname{}}. 
A small $\lambda$ \revise{would }{}highlight\revise{}{s} \revise{even }{}small differences which advanced dancers may use to achieve perfect synchronization.
A large $\lambda$ \revise{can be}{is} useful to suppress such minor discrepancies and only highlight major differences.

\subsubsection{Overall Pose Similarity}
Given the $BPD(i,t)$ values for all body parts \revise{at}{in} a frame, we then compute an \textit{Overall Pose Similarity} ($OPS(t)$) value. 
We \revise{decided to use}{used} a Support Vector Regression (SVR) model~\cite{svr} as our study found that it was the most robust (\revise{please refer to our system evaluation presented in}{see} Section~\ref{sec: comparison}).
In our current implementation, our model takes all 13 $BPD(i,t)$ values as \revise{the }{}input\revise{,}{} and predicts a\revise{}{n} $OPS(t)$ value that ranges from 0 (totally ill-synchronized) to 1 (perfectly synchronized).
The system uses the predicted $OPS(t)$ values \revise{in}{for} the 1D heatmap (Figure~\ref{fig: interface_pose}).

\subsection{Temporal Alignment Analysis}

The temporal alignment of movements is another aspect of synchronization \revise{we consider}{considered} in SyncUp.
We extend Davis and Agrawala's~\cite{Davis2018VisualBeat} notion of \revise{a }{}directogram to quantify motion rhythms.
% in a video for synthesizing dancing-like motions.
This notion is similar to a spectrogram, which factors volume changes into frequencies, except that a directogram factors motions \revise{in}{of} a video into different angles.
We further modify their method to compute \revise{a }{}directogram-equivalent metric\revise{}{s} for body parts of multiple dancers.
% We first define the pose flow of the $i$-th keypoint of the $j$-th dancer ($\vec{f_j}(i,t)$) as the vector difference of the :

We first define the \textit{pose flow} of one dancer $\vec{f_j}(t)$, which describes the movement of \revise{the }{}dancer $j$ at \revise{the }{}frame $t$:

\begin{equation}
    \vec{f_j}(t) = \{p_j(i,t+1) - p_j(i,t)\}, i \in \{1, 2,..., 18\}
    \label{equ: kp_diff}
\end{equation}

where $p_j(i,t)$ represents the $i$-th AlphaPose keypoint at \revise{the }{}frame $t$.
We use \revise{the }{}all 18 AlphaPose keypoints to compute $\vec{f_j}(t)$ in our current implementation.

We then compute a \pgram ($\mathcal{P}(t,\theta)$) for each dancer in a similar manner to the directogram:
\begin{equation}
    \mathcal{P}_{j}(t,\theta) = \sum_{i} |\vec{f_j}(t)|  \mathds{1}_\theta(\angle \vec{f_j}(t)),
    \hspace{0.5cm}
 \mathds{1}_\theta(\phi):=   \left\{\begin{array}{ll}      
1, & \textnormal{if } |\theta - \phi| \leq \frac{2\pi}{N_{bins}} \\      
0, & \textnormal{otherwise}\\
\end{array} \right.
\end{equation}

The \pgram factors body\revise{ }{-}part motions of a dancer in\revise{}{to} $N_{bins}$ different angles.
For example, when a dancer \revise{is stretching}{stretches} only the right hand horizontally, \revise{her}{the} \pgram \revise{would}{will} show a large value in the corresponding direction while the other values \revise{would be}{are} nearly zero.

\begin{figure}[t]
    \centering
    % \includesvg[width=\linewidth]{figures/Quantification/Tempo_explain.svg}
    \includegraphics[width=\linewidth]{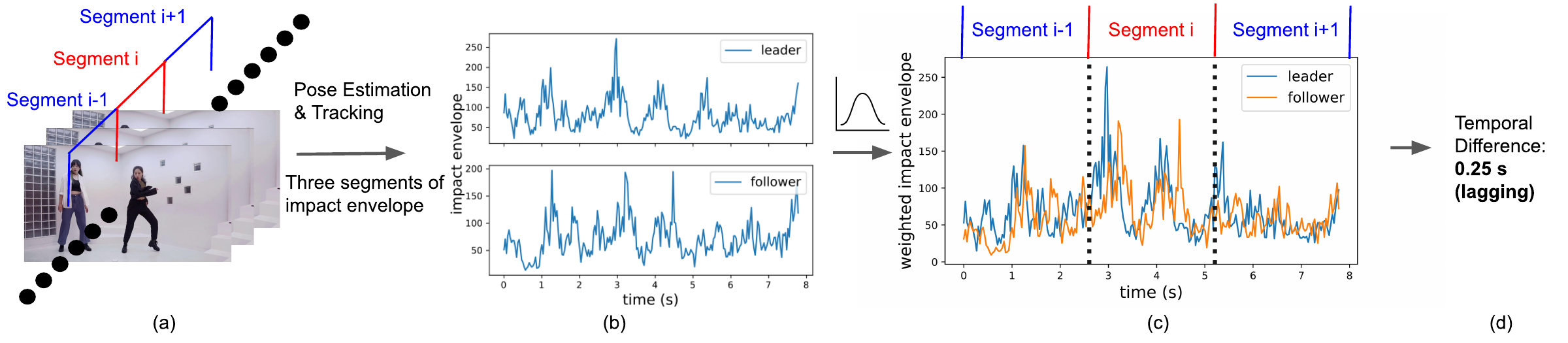}
    \caption{The Workflow of our temporal alignment analysis.
    (a)->(b): \systemname{} first computes human poses and performs dancer tracking. 
    (b)->(c): To analyze the temporal alignment at the video segment $s$, \systemname{} samples impact envelopes (Equ. \ref{equ: impact}) at the segment $s-1$, $s$, and $s+1$.
    The system then applies a Gaussian weight distribution on impact envelopes over these three segments to weigh the contributions by the segment $s$.
    The system calculates the impact envelopes for the leader dancer and one follower.
    (c)->(d): The system then computes the cross-correlation between the two envelopes. In this example, the calculation result is 0.25 second, suggesting that the follower's movements are behind those of the leader.
    }
    \label{fig:imp}
\end{figure}

Similar to Davis and Agarawala's work, we \revise{can }{}further define a flux $\mathcal{P}_{F, j}(t,\theta)$ and an impact envelope $u_{j}(t)$ of a \pgram as follows:
\begin{equation}
    \mathcal{P}_{F, j}(t,\theta) = \mathcal{P}_{j}(t,\theta) - \mathcal{P}_{j}(t-1,\theta),
    \hspace{0.5cm}
    u_j(t) = \sum_{\theta=\theta_i}^{\theta_{N_{bins}}}  |\mathcal{P}_{F, j}(t,\theta) |
    \label{equ: impact}
\end{equation}
The flux is the temporal differentiation of \revise{the }{}consecutive \pgrams, which is \revise{an analogy}{analogous} to acceleration. 
The impact envelope sums up all \revise{the flux}{fluxes} within a dancer along the angular bin axes.
This highlights how much one person's motion \revise{is changing}{changes} at one time.
Figure~\ref{fig:imp}b shows an example of impact envelopes of two different dancers. 

The impact envelopes \revise{serve to identify}{identifies} the temporal differences of visual beats between two dancers.
% .\zhongyi{Should we clarify what is visual beat? though it is defined in their work.}
We use three consecutive segments to secure a sufficient number of samples for reliable calculation of cross-correlation.
We also apply Gaussian weights to impact envelopes in three consecutive segments
%(each segment is 8-beat long as we explained above)
to highlight the degree of the temporal alignment at the center segment.
The Gaussian weights also help \revise{to }{}reduce artificial noise caused by the cutoff of the impact envelopes.
Cross-correlation analysis identifies how much time shift \revise{would give}{will offer} the best match between the two impact envelopes.
The time shift \revise{that exhibits}{exhibiting} the highest cross-correlation value is considered as the degree of the temporal alignment against the reference dance, denoted as $\tau_{j}(s)$ ($s$ is a segment).
\systemname{} uses the accumulation of the absolute values of the temporal alignment scores calculated between all leader-follower pairs (i.e., $\tau(s) = \sum_j |\tau_{j}(s)|$) in the 1D heatmap visualization (Figure~\ref{fig: interface_temp}).

Figure \ref{fig:imp} summarizes the workflow of how \systemname{} computes the temporal alignment at each segment.
Figure~\ref{fig:imp}c shows two examples of impact envelope plots after the Gaussian weights are applied.
The dotted lines are the boundaries of the segments.
By looking at the peaks and valleys of the two plots, the figure suggests that the follower's dance (orange line) is behind the leader's dance (blue line).
$\tau_{j}(s)$ is 0.25 \revise{second}{s} in this example, confirming the delay.

%% file: contents/eval.tex
\section{System Evaluation}
We conducted system evaluations on the two dance performance quantification methods.
For our evaluations \revise{as well as}{and the} training of the neural networks for pose similarity analysis, we created a collection of dance videos\revise{,}{} and \revise{performed labelling of the degree of synchronization by human raters}{the corresponding labels of the degree of synchronization}. 

\subsection{Dance Video Data Collection}
Because we have two practice modes (individual and group\revise{ practice modes}{}), we collected two types of dance videos for our evaluations as well as training of learning-based approaches.

\begin{figure}
    \centering
    % \includesvg[width=\linewidth]{figures/SystemEval/IPV_Data.svg}
    \includegraphics[width=\linewidth]{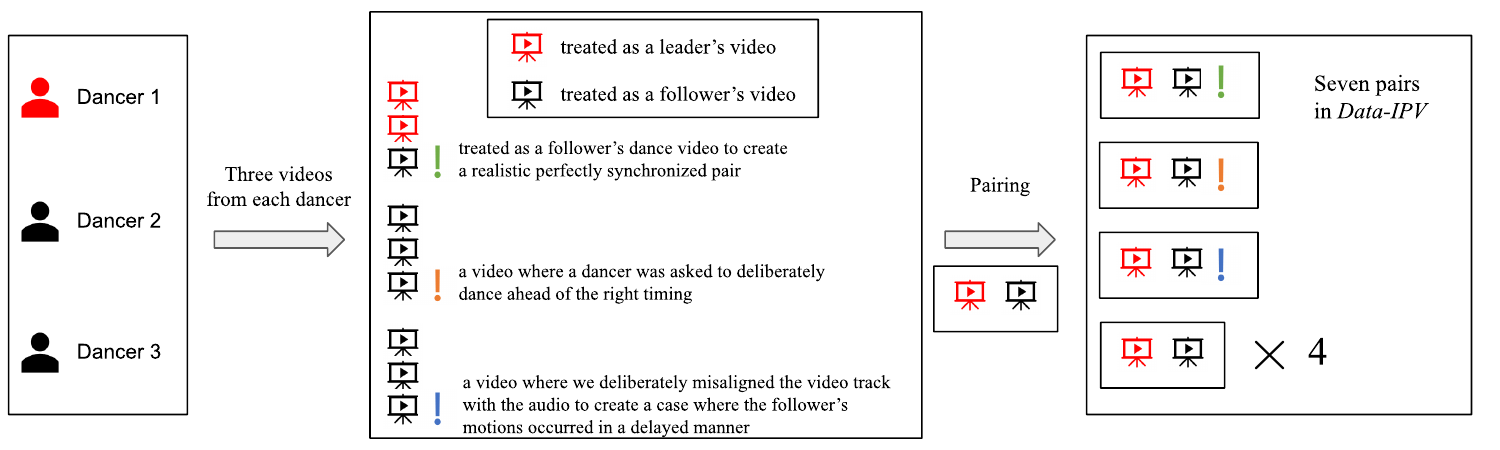}
    \caption{Composition of \textit{Data-IPV}.
    Each dancer in one group provided 3 practice videos, resulting 9 videos in total. 
    Two of the 3 leader's videos were then treated as the references  (\iconvideored). To cover different degrees of synchronization, we deliberately created pairs to represent extreme cases where synchronization was close to or far from perfection (i.e., pairs with \iconvideoblack with exclamation marks). In this manner, we created seven pairs from each dance group.}
    \label{fig: ipv_pair}
\end{figure}

\subsubsection{Individual Practice Video Pairs (\textit{Data-IPV})} 

We collected videos where individual dancers \revise{were practicing}{practiced} separately.
We recruited one volunteer group, consisting of three dancers, for this part of data collection.
All of them practiced synchronized dancing that Asian pop stars or anime characters originally perform\revise{}{ed}.
We asked each volunteer to perform practices of three dance routines individually and capture their practices \revise{with}{using} their smartphones or \revise{W}{w}ebcams.
This led to \revise{9}{nine} videos from one group, and we used them to create \revise{7}{seven} pairs.

Figure \ref{fig: ipv_pair} shows how we created these pairs.
Our objective\revise{ of this pairing design is}{was} to create a wide range of in-\revise{synchronization}{} and out-of-synchronization samples from the limited \revise{amount}{number} of dance data \revise{we }{}collected.
To create a realistic perfectly-synchronized pair, we paired two videos from the leader (the pair that included \iconvideogreen~in Figure~\ref{fig: ipv_pair}).
When collecting the dancing data, we asked one follower to deliberately dance ahead of the \revise{right}{correct} timing (\iconvideoorange~in Figure~\ref{fig: ipv_pair}). 
By pairing this with the reference video, we created a pair that represented a case where the follower's motions occurred ahead of those of the leader.
Furthermore, to create a pair that represents a case where the follower’s motions occurred in a delayed manner, we deliberately introduced a delay of 100 \revise{milliseconds}{ms} only to the video footage (but not the audio track) in \revise{one of the }{an}other follower's practice videos (\iconvideoblue~in Figure~\ref{fig: ipv_pair}). 
We confirmed that \revise{it}{this} caused sufficient lagging effects throughout the video for our ML approaches.
These three pairs were created to cover extreme cases where synchronization was close to or far from perfection.
We then paired the remaining four videos (\iconvideoblack without the exclamation mark) to the reference videos (\iconvideored).
As a result, we created seven pairs of videos from each dance group.

\subsubsection{Group Practice Videos (\textit{Data-GPV})}
To evaluate our dance performance quantification methods in the scenario of group practices, we also gathered videos where multiple dancers \revise{were dancing in the same video}{danced together}.
We recruited another four dance groups (2\revise{ -- }{--}4 dancers in each group) and received \revise{9}{nine} videos of group practices from them in total.
Their dance style was similar to that of the individual practice video data.
\begin{figure}[t]
    \centering    \includegraphics[width=0.9\linewidth]{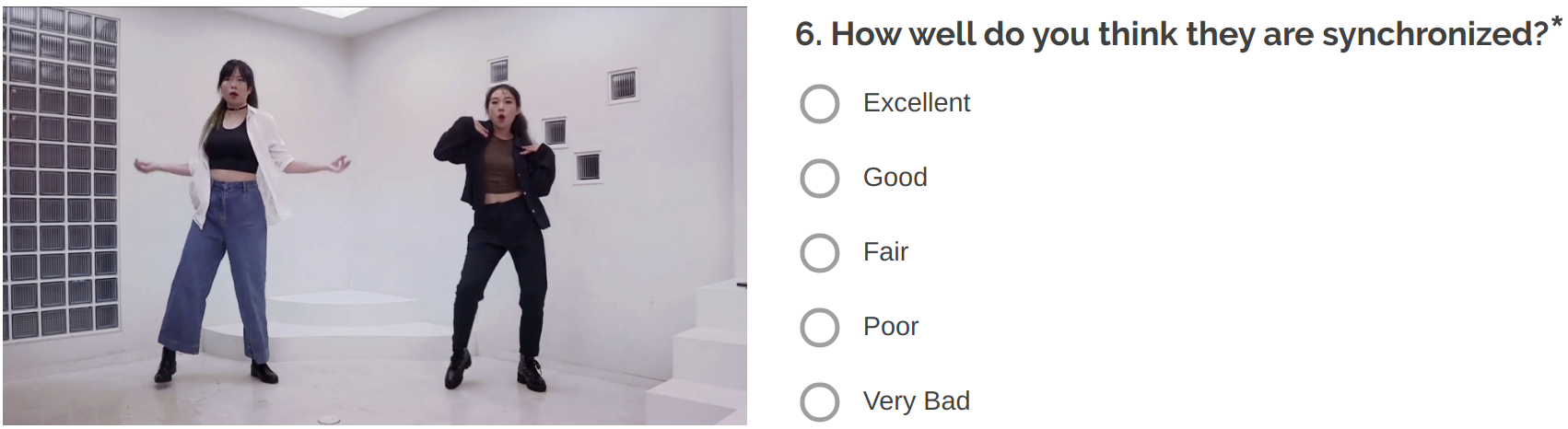}
    \caption{
    \revise{The W}{A w}eb interface \revise{}{was }used for collecting human ratings on pose similarity. We used a similar system for human assessment collection on the temporal alignment evaluation.}
    % The upper frame is a sample from remote-practice mode, in which we combine two frames from two separate videos by cutting out the central parts.
    \label{fig: eval_survey_web}
\end{figure}

\subsection{Dance Performance Assessments with Human Raters}
\subsubsection{Pose Similarity}
\revise{As}{Because} the pose similarity algorithm uses frames of a dance video, we created a set of frames and asked our volunteers to rate how well the poses of the two dancers were synchronized\revise{ in the given frames}{}.
We randomly-sampled 100 frames from each of the two data source\revise{}{s} (\revise{}{i.e., }\textit{Data-IPV} and \textit{Data-GPV}), resulting in 200 frames in total.
The number of random samples \revise{we chose}{chosen} from each practice video was set to be proportional to its video length.
We also confirmed that the seed for each data source was \revise{was }{}set to be different in our random sampling so that we did not include \revise{the }{}frames of \revise{the }{}identical \revise{portion of a dance routine}{dance portions} in both training and testing data.
We then created an online form
where our volunteers shared their perceived pose similarity of the two dancers with \revise{a 5-Likert scale response}{according to a 5-Point Likert scale} (``Excellent'', ``Good'', ``Fair'', ``Poor'', and ``Very Bad''), which was internally mapped to \revise{the }{}numerical scores of 1, 0.75, 0.5, 0.25, and 0, respectively.

We recruited 161 volunteers (98 males, 60 females, and 3 prefer not to say) \revise{in total }{}for this part of the rating collection.
Each volunteer was assigned to randomly\revise{-}{ }selected 25 frames.
As a result, we obtained 20 ratings on average for each frame.
We removed outliers that were beyond \revise{3}{three} standard deviations from the mean rating \revise{in}{of} each sample, which is a common heuristic for outlier removal.
Our algorithm regards the mean of the remaining ratings as the overall human rating of the pose similarity.
We treated this overall human rating as a continuous value and used it as the training label for our following learning-based \revise{methods}{regression models}.

\subsubsection{Temporal Alignment}
For evaluating the temporal alignment algorithm, we extracted 20 frames (randomly-selected 10 scenes from each of \textit{Data-IPV} and \textit{Data-GPV}).
We then created another online form similar to that for pose similarity for collecting ratings from our volunteers.
To clarify our focus in this part of the rating collection, we explicitly instructed our volunteers to judge only the temporal alignment of two dancers in the given frame.
We deliberately slowed down the playback speed of each frame (\revise{at 70\% of the original speed}{70\% of original}) so that the volunteers could closely examine the temporal alignment.
We also clarified which dancer was the reference in the instruction.
The volunteers were asked to rate their perceived temporal alignment \revise{with a 5-Likert scale response}{according to a 5-Point Likert scale} (``A lot faster'', ``A little faster'', ``At the same time", ``A little slower'', and ``A lot slower''), which corresponded to the numerical scores of 1, 0.5, 0, -0.5, and -1, respectively.

We recruited 15 volunteers (\revise{8}{eight} males and \revise{7}{seven} females) for this rating collection.
All volunteers rated the whole set of \revise{the }{}20 frames.
We employed the same outlier exclusion as in the pose similarity, but we did not find any outlier.
Therefore, we directly used the mean value of the ratings for each frame as the ground truth temporal alignment score.

\subsection{Comparison for Pose Similarity}
\label{sec: comparison}
\revise{In order to}{To} evaluate both the accuracy and robustness of our prediction of $OPS$, we experimented four different methods\revise{}{,} including the SVR-based approach used in our current SyncUp implementation.

\subsubsection{Simple Addition}
The most straightforward approach to comput\revise{e}{ing} $OPS(t)$ \revise{is based on a}{uses the} na\"{i}ve sum of all \revise{the }{}$BPD(i,t)$ values at every body part.
We first normalized the sum of $BPD(i,t)$ values into 
[0, 1] (denoted as $\overline{S_{BPD}(i,t)}$).
To match \revise{with }{}our human ratings, we applied the reverse conversion on $\overline{S_{BPD}(i,t)}$ (i.e., $OPS(t) = 1 - \overline{S_{BPD}(i,t)}$).
The method based on the simple addition is commonly used in existing research~\cite{ chen2010learning, zhou2019visualizing}.
One underlying assumption is that all \revise{the }{}features (in our case, $BPD$) should weigh equally for the final overall score.
This assumption does not necessarily fit human perception of pose similarity~\cite{chen2010learning, muller2005efficient}.
We used this as a baseline method in our evaluation.

\subsubsection{Neural Network}
Neural networks (NNs) are often appropriate for complicated perceptual tasks\revise{ like}{, such as} object detection and face recognition~\cite{Girshick2014Girshick, Alex2012ImageNet}.
We included two types of NNs in our performance comparison: short-NN and long-NN.
Short-NN consisted of two layers: \revise{the}{an} input layer (i.e., $BPS$ with 13 dimensions) and \revise{the}{an} output layer ($OPS$, one dimension).
Long-NN consisted of four layers: \revise{the}{} input \revise{layer}{}, two hidden \revise{layers}{} (10-dimension and 5-dimension) and \revise{the}{} output \revise{layer}{}.
More complicated network structures may outperform, but we decided to experiment rather simple networks due to the limited number of samples for training.
We used $LeakyReLU$ as the activation function after each linear transformation.
For training, we used the \revise{RMSE (Root Mean Squared Error)}{Root Mean Squared Error (RMSE)} loss function and Adam~\cite{kingma2014adam} as the optimizer with the learning rate of $0.01$.
The whole training session took 50 epochs.

\subsection{Results}

\begin{table}[t]\centering
  \small
\caption{\revise{A c}{C}omparison of OPS prediction results. The results are shown in the format of ``RMSE (Pearson's $r$ value)''. All \revise{the }{}$p$ values for Pearson's $r$ \revise{}{values }were smaller than .05.}
\label{tab: ops_overall_acc}
% \scriptsize
\begin{tabular}{lrrrrrrrrrrr}
\toprule
\multirow{2}{*}{\textbf{Method}} &\multicolumn{10}{c}{ $\lambda$} \\\cmidrule{2-11} &
% \cmidrule{1-10}
\textbf{0.333} &\textbf{0.426} &\textbf{0.543} &\textbf{0.693} &\textbf{0.885} &\textbf{1.13} &\textbf{1.44} &\textbf{1.84} &\textbf{2.35} &\textbf{3} \\\midrule
\textbf{Simple Addition} & \shortstack{0.168\\(0.65)} &\shortstack{0.170\\(0.67)} &\shortstack{0.174\\(0.69)} &\shortstack{0.181\\(0.71)} &\shortstack{0.191\\(0.73)} &\shortstack{0.205\\(0.72)} &\shortstack{0.237\\(0.69)} &\shortstack{0.278\\(0.64)} &\shortstack{0.317\\(0.64)} &\shortstack{0.348\\(0.59)} \\ \midrule
\textbf{SVR} & \shortstack{0.163\\(0.63)} &\shortstack{0.159\\(0.65)} &\shortstack{0.156\\(0.66)} &\shortstack{0.154\\(0.68)} &\textbf{\shortstack{0.151\\(0.69)}} &\shortstack{0.153\\(0.69)} &\shortstack{0.158\\(0.66)} &\shortstack{0.161\\(0.64)} &\shortstack{0.169\\(0.59)} &\shortstack{0.180\\(0.51)} \\ \midrule
\textbf{Short\_NN} & \shortstack{0.204\\(0.23)} &\shortstack{0.216\\(0.14)} &\shortstack{0.199\\(0.34)} &\shortstack{0.174\\(0.55)} &\shortstack{0.165\\(0.61)} &\shortstack{0.164\\(0.62)} &\shortstack{0.170\\(0.58)} &\shortstack{0.177\\(0.55)} &\shortstack{0.184\\(0.49)} &\shortstack{0.190\\(0.45)} \\ \midrule
\textbf{Long\_NN} & \shortstack{0.172\\(0.57)} &\shortstack{0.173\\(0.60)} &\shortstack{0.166\\(0.61)} &\shortstack{0.167\\(0.61)} &\shortstack{0.167\\(0.61)} &\shortstack{0.165\\(0.62)} &\shortstack{0.170\\(0.61)} &\shortstack{0.170\\(0.60)} &\shortstack{0.175\\(0.58)} &\shortstack{0.176\\(0.54)} \\
\bottomrule
\end{tabular}
\end{table}

\begin{table}[t]
    \centering
    \small
    \caption{The prediction results with \textit{Data-GPV} and \textit{Data-IPV}.
    The results are shown in the format of ``RMSE (Pearson's $r$ value)''.
    We highlight the lowest RMSE in each dataset condition.
    All $p$ values were smaller than .001.}
    \begin{tabular}{cccc}
    & \textit{Data-GPV} & \textit{Data-IPV} & Both \\
    \hline
    Addition & 0.193 (0.54) & 0.189 (0.88) & 0.191 (0.73) \\
    SVR & \textbf{0.174 (0.45)} & \textbf{0.124 (0.87)} & \textbf{0.151 (0.69)} \\
    Short\_NN & 0.180 (0.34) & 0.149 (0.80) & 0.165 (0.61) \\
    Long\_NN & 0.193 (0.33) & 0.137 (0.82) & 0.167 (0.61) \\
    \end{tabular}
    % \caption{Caption}
    \label{tab: datasetcompare}
\end{table}

\subsubsection{Pose Similarity}
We performed prediction\revise{}{s} using the \revise{the }{}four $OPS$ computation methods (i.e., Simple Addition, SVR, Short-NN, and Long-NN) with different $\lambda$ values. 
We chose 10 values uniformly distributed between 0.333 and 3 in a logarithmic scale.
An overly low or high $\lambda$ was excluded \revise{as}{because} it was not practical (i.e., too many differences or none \revise{of them }{}were highlighted in the visualization, respectively).
We employed cross validation where one of the 16 data sources \revise{that were }{}used for creating static frames (Section 6.2.1) was reserved for testing, and the remaining data were used for training to evaluate our SVR and NN-based approaches. This led to 16-fold cross validation (\revise{7}{seven} pairs from \textit{Data-IPV} and \revise{9}{nine} dances from \textit{Data-GPV}).

\begin{figure}[t]
    \centering
    \begin{subfigure}[b]{0.23\linewidth} 
        \centering
        \includegraphics[width=\textwidth]{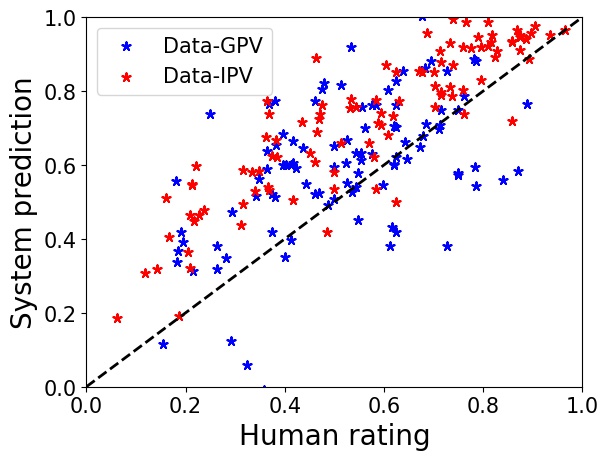}
        \caption{Simple Addition}
        \label{fig: dot_add}
    \end{subfigure}
    % \\
    \begin{subfigure}[b]{0.23\linewidth}
        \centering
        \includegraphics[width=\textwidth]{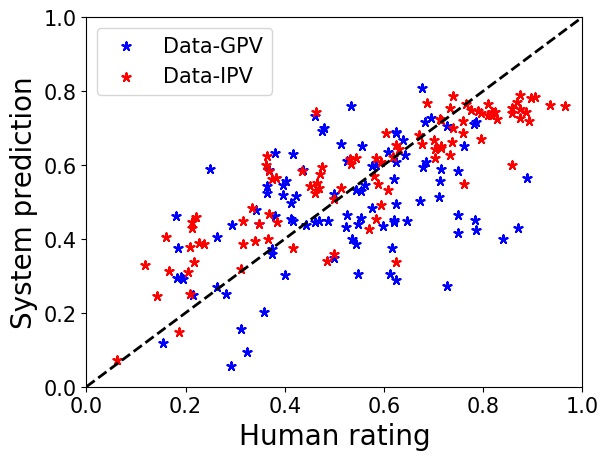}
        \caption{SVR}
        \label{fig: dot_SVM}
    \end{subfigure}
    \begin{subfigure}[b]{0.23\linewidth}
        \centering
        \includegraphics[width=\textwidth]{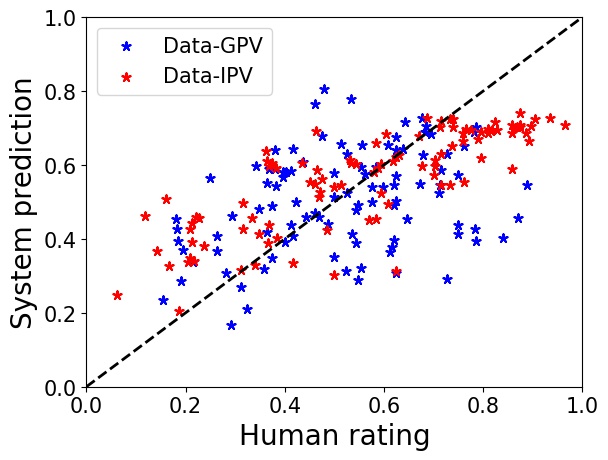}
        \caption{Short Neural Network}
    \end{subfigure}
    \begin{subfigure}[b]{0.23\linewidth}
        \centering
        \includegraphics[width=\textwidth]{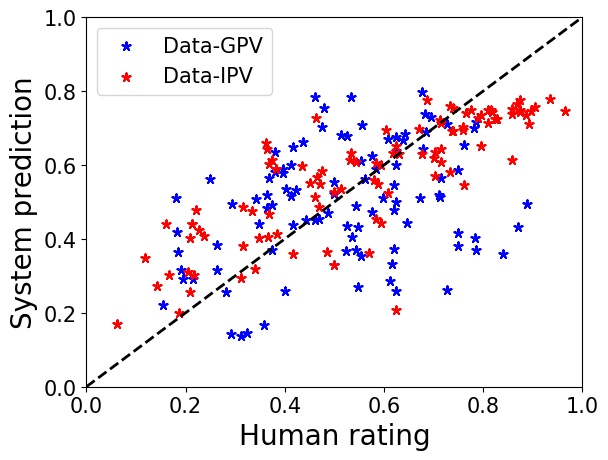}
        \caption{Long Neural Network}
    \end{subfigure}
    \caption{Scatter plots of system predictions and average human ratings for the four OPS computation methods. $\lambda$ was set to 0.885 in these results. The black dotted line represents the perfect prediction.}
    \label{fig: dotmap} 
\end{figure}

Table~\ref{tab: ops_overall_acc} shows the \revise{root mean squared error (RMSE)}{RMSE} and Pearson's $r$ values between the human rating averages and predictions by the four methods under \revise{the }{}all $\lambda$ conditions. 
The $p$ values for \revise{the }{}all $r$ were smaller than .05.
With all $\lambda$ values, we found that the SVR-based method demonstrated the smallest RMSE.
This suggests that the SVR-based method was the most accurate and robust in our examination.
We also found that correlation values of Simple Addition and SVR were close while the values of the other two methods were lower.
%\subsubsection{Prediction distribution of the samples}
Figure~\ref{fig: dotmap} further visualizes the detailed results of all \revise{the }{}200 samples with $\lambda=0.885$.
The plots compare their human ratings (\revise{the }{}x-axis) and predictions (\revise{the }{}y-axis).
The red and blue dots represent the results of \textit{Data-IPV} and \textit{Data-GPV}, respectively.
We found that the red samples tended to be positioned more closely to the black dotted line (\revise{the }{}perfect prediction).
Table~\ref{tab: datasetcompare} summarizes the accuracy of different methods using \textit{Data-GPV}, \textit{Data-IPV}, and both under the condition of $\lambda=0.885$. 
The results show that the overall accuracy in \textit{Data-IPV} was higher than that in \textit{Data-GPV}.

\subsubsection{Temporal Alignment}
\begin{figure}[t]
\centering
\begin{minipage}[t]{.47\textwidth}
    \centering
    \includegraphics[width=\linewidth]{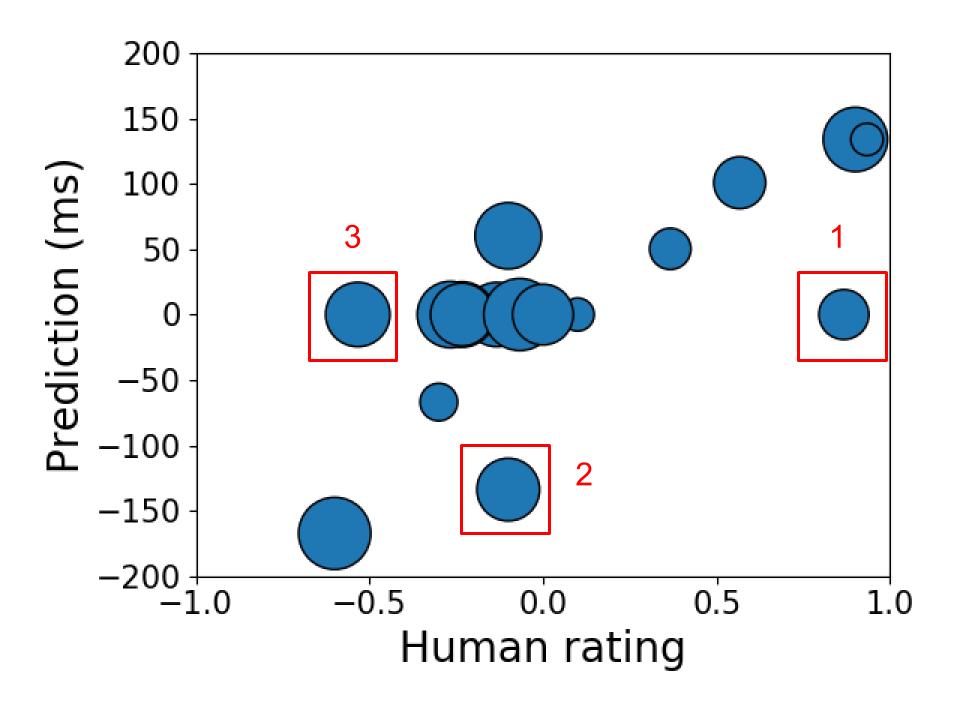}
    \captionof{figure}{A scatter plot for the human ratings and \systemname{} temporal alignment predictions. We highlight three cases with a red rectangle where our predictions were largely different from the ground truth. The number annotations are included for the discussions in the paper.
    }
    \label{fig:eval_temp} 
\end{minipage}%
\hfill
\begin{minipage}[t]{.47\textwidth}
      \centering
      \includegraphics[width=\linewidth]{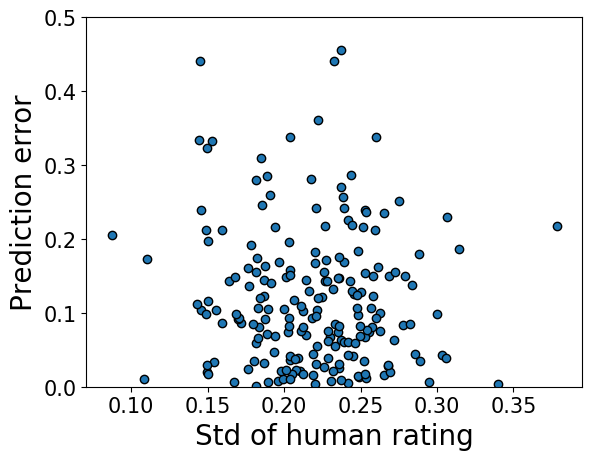}
      \captionof{figure}{A scatter plot \revise{between}{of} the standard deviations \revise{of}{vs.} human ratings and absolute prediction errors. The Pearson's $r$ \revise{}{value }was -.08 ($p=.26$).}
      \label{fig:eval_err} 
\end{minipage}
\end{figure}

To calculate the temporal alignment with our algorithm, we extracted the previous and next segments of each segment given to the raters from the original video data.
Figure~\ref{fig:eval_temp} presents the comparison of the temporal alignment results between human ratings and our prediction.
Note that human ratings \revise{we}{were} collected \revise{were in a 5-Likert scale}{according to a 5-Point Likert scale} (from -1 to 1)\revise{}{,} whereas our prediction was given \revise{in an}{using the} estimated time differen\revise{t}{ce} (ms\revise{ec}{}).
There are seven samples within \revise{the}{a} region of ($\pm$0.2, $\pm$20 [ms\revise{ec}{}]).

The Pearson's $r$ was .712 ($p<.001$), showing a strong positive correlation.
In the two cases where raters unanimously marked \revise{as }{}``at the same time (\revise{the}{a} score of \revise{0}{zero})'', our predictions \revise{are}{were} both\revise{ equal to}{} zero.
There are three notable cases where our predicted scores were quite far from the human ratings (highlighted with red squares in Figure~\ref{fig:eval_temp}).
Excluding the three highlighted cases, \revise{all }{}the other 17 samples in Figure~\ref{fig:eval_temp} shows that \systemname{}'s temporal alignment predictions were highly correlated with human ratings (\revise{the }{}Pearson's $r=$ .890 ($p<.001$)).

\subsection{Discussion}

\subsubsection{Pose Similarity}

The evaluation results of pose similarity demonstrate a smaller RMSE by our prediction (0.151) than human raters (0.223).
The results suggest that our quantification method \revise{could}{can} predict a closer score of pose similarity to the ground\revise{ }{-}truth data (i.e., the average of human ratings) than raters.
However, it does not necessarily mean that our method is more accurate than human raters in general.
Quantifying pose similarity is an inherently challenging task as even human raters demonstrate\revise{d a}{} certain degree\revise{}{s} of disagreement.
In one case, the standard deviation of human ratings was as large as \revise{0}{}.378.
Human raters may have weighed different aspects of dancing, potentially leading to \revise{such }{}a diverse set of ratings.
Future work should investigate what caused \revise{}{the }large disagreement\revise{}{s} in some cases\revise{,}{} and how algorithms \revise{could take}{might account for} such disagreement\revise{ into account}{}.

The Pearson's $r$ values in the Simple Addition method were slightly higher than those in the SVR-based method.
\revise{It}{This} implies that the data distribution in the Simple Addition method may demonstrate linearity along\revise{ with}{side} a different slope and intercept.
Learning-based models can be expected to accommodate such biases, and the SVR-based method and NNs offer linear and non-linear mapping, respectively.

Our results show\revise{ed}{} that the SVR-based method performed well over \revise{neural networks}{NNs}.
Neural network methods usually perform well when they are applied directly to raw data, (e.g., images and texts).
They can extract high-level semantic features within these data.
However, in our cases, the input of our ML models was a set of manually\revise{-}{ }crafted feature vectors because we were not able to have a sufficient \revise{amount}{number} of labeled data for training \revise{neural networks}{NNs} with raw data.
\revise{A}{Furthermore, a}ppropriate pre-trained models were not available for our purpose\revise{ either}{}.
This configuration \revise{might}{may} have impacted the performance of the two methods using \revise{neural networks}{NNs}.
Another possible explanation for the results in favor of the SVR-based method is that the intrinsic function that maps \revise{between }{}the input vectors \revise{and}{to} expected output values was not highly complex unlike image recognition and text translation.
Future work is encouraged to further examine the capability of various machine learning approaches, but based on our RMSE results across different $\lambda$ values (Table~\ref{tab: ops_overall_acc}), we decided to employ the SVR-based method in our current SyncUp prototype.

One major cause of the lower accuracy in \textit{Data-GPV} was \revise{}{the }occlusion of body parts (\revise{please }{}refer to Appendix~\ref{appendix} for more details).
Similar to other computer vision methods, occlusion \revise{would lead}{leads} to less reliable pose detection\revise{}{s}.
We closely examined all the segments in \textit{Data-GPV}\revise{,}{} and found that 14 \revise{samples }{}included occlusion\revise{}{s} of a dancer\revise{ by the other}{}.
When excluding these 14 samples, we found the RMSE of \textit{Data-GPV} improve\revise{s}{d} to 0.174 from 0.146.

We examined the relationship between \revise{the }{}the standard deviation of human ratings and prediction errors.
We expected that prediction \revise{c}{w}ould be less precise in cases where human ratings diverged.
Figure~\ref{fig:eval_err} presents the results \revise{with}{vs.} our SVR-based approach with $\lambda=0.885$ (showing the best performance in Table~\ref{tab: ops_overall_acc}).
The Pearson's $r$ \revise{}{value }was -.08 ($p=.26$).
Therefore, we were not able to confirm a clear connection between the degree of rater\revise{s'}{} agreement and prediction accuracy.
This result \revise{also suggests}{reflects} the challenging nature of predicting pose similarity using computer vision approaches\revise{}{,} though our SVR-based method performs well in general.

\subsubsection{Temporal Alignment}

Our results showed that the predictions of temporal alignments \revise{were }{}correlated well with human ratings.
We closely examined the three cases where we observed large errors.
In the following section, we discuss these cases in detail and each case corresponds to \revise{the }{}annotation\revise{}{s} in Figure~\ref{fig:eval_temp}. 
\begin{itemize}
%  \item The individual entries are indicated with a black dot, a so-called bullet.
%  \item The text in the entries may be of any length.
\item Case 1.
In this case, the average human rating was \revise{0}{}.87, suggesting that our raters considered that the movements of a follower in the video \revise{was}{were} largely delayed.
However, the \systemname{} prediction was 0 [ms\revise{ec}{}].
By inspecting the corresponding video, we found
that the follower skipped some movements to catch up with the lead\revise{er}{} dancer.
Although our method does not consider such skips, the human raters took them into account for the rating, resulting in a large discrepancy in the temporal alignment score.
However, our pose similarity analysis can identify such cases, and thus this observation confirms that \systemname{} should present both pose similarity and temporal alignment results to users.

\item Case 2. 
In this case, the average human rating was -\revise{0}{}.10, but the \systemname{} prediction was -134 [ms\revise{ec}{}].
We also confirmed that the dancers were well synchronized in the given segment.
However, the follower dancer was largely ahead of the leader in the two adjacent segments which were not shown to our raters.
We thus concluded that the performance in these two adjacent segments affected our temporal alignment prediction.

\item Case 3.
In this case, the average human rating was -\revise{0}{}.53, but the \systemname{} prediction was 0 [ms\revise{ec}{}].
In the corresponding video, we found that the poses of the two dancers were largely different although the timing of their movements was the same.
We contacted the dancers who provided this video footage and they agreed that they were synchronized in terms of the temporal alignment by commenting: \textit{``We truly look different, but the movements are on-beat together.''}
We thus concluded that our prediction was correct, but the large pose difference between the dancers \revise{had }{}caused a bias among the human raters.
This case suggests a benefit of \systemname{}, \revise{}{in }which \revise{can separate }{}the contributions \revise{by}{of} poses and movement timing \revise{}{can be distinguished }to accurately identify segments that exhibit a low degree of synchronization.

\end{itemize}

%% file: contents/userstudy.tex
\section{Qualitative User Evaluation}

We next conducted a user study to understand the potential benefits of SyncUp in actual use.
Due to the spread of COVID-19, we were \revise{not able}{unable} to \revise{run}{conduct} an in-person user study.
We instead conducted an online qualitative study to obtain feedback from our target users.

We recruited \revise{3}{three} new dance groups (\revise{9}{nine} dancers in total, \revise{8}{eight} females and \revise{1}{one} male, ages: 24--32). 
All dancers were amateur synchronized dancers practicing k-pop dances.
We asked each group to perform the same dance routine (65--78 \revise{seconds}{s} long) 12 times and \revise{shared}{to share} the recordings with us.
We then invited the leader dancers of these three groups (DL 1--3) for our qualitative study.
We first presented \revise{them}{} our SyncUp interface (including the spotlight view) \revise{with}{alongside} the videos they shared with us.
After explaining \revise{all}{} the features, they were invited to use our system.
To encourage active interaction, we asked them to perform tasks of identifying segments where dancers were not well synchronized (ill-synchronized segments)  with our system as well as a standard Web video player interface.
We also asked \revise{our participants}{them} to experience the spotlight view and perform the same task.
After they agreed that they understood the functionalities of SyncUp, we conducted a semi-structured interview that included questions about how the system could be useful in their actual practices and how it could change interactions and communication\revise{}{s} among group members.
\revise{At the end of the study}{After the study}, each participant, including those who did not attend our interviews, was offered approximately \revise{70 USD}{USD 70} in \revise{a}{their} local currency as compensation.

\revise{In addition, we}{We also} recruited another six dancer pairs (23--35 years old, and 11 female and \revise{1}{one} male) only for the interview part.
The experimental procedure was the same, except that we used the videos offered by the three groups \revise{above}{prior} this \revise{time}{interview}.
\revise{In this manner}{Hence}, we reduced the participants' burden for participation.
We recruited the participants as pairs so that they had opportunities to discuss their thoughts with each other during our interviews.
We refer to the dancer pairs as DP\revise{s}{} 1--6.
They were offered approximately 40 USD in a local currency as compensation at the completion of the study.

We transcribed all \revise{the }{}interviews and analyzed \revise{}{the information} by categorizing quotes \revise{with}{using} the open coding approach~\cite{moghaddam2006coding}.
Open coding is a qualitative analysis approach by categorizing observed quotes that describe the same or similar phenomena and creating names (or codes) that represent the corresponding categories.
\revise{As}{Because} none of our participants (\revise{both }{}leaders \revise{and}{or} pairs) was fluent in English, we conducted interviews in their \revise{local}{spoken} language.
We translated the quotes as faithfully as possible for the report in this paper.

\subsection{Results}

\subsubsection{Enabling Quick Access to Ill-Synchronized Segments}
All participants appreciated the design of visualizations in \systemname{}.
DP2 and DP6 explicitly rementioned the inefficiency of dragging the seek bar of a video player for locating ill-synchronized segments.
On the other hand, \revise{the participants}{they} saw clear values in \revise{}{the} visualizations \revise{SyncUp offers}{offered by SyncUp}.

\myquote{``These (1D heatmaps) are intuitive. It helps me quickly locate these [ill-synchronized] segments, and I do not need to watch all the videos.''}{DP1}

\myquote{``This heatmap [overlays] is very clear. I can easily know where the problems are. Very useful.''}{DP2}

\myquote{
``I like those vertical comparisons [in 1D heatmaps]. I can clearly know in which one (segment) dancers are not well synchronized in each practice. We can then focus on practicing these parts (segments).''}{DL-3}

Our participants also enjoyed the spotlight view and commented its potential benefits.

\myquote{``I really like this function. It can intelligently recommend us segments in which we are very synchronized or out of synchronization.''}{DP5}

\subsubsection{Offering an Objective Assessment on Dance Performances}
In five of the interviews (DL2, DP1, DP3, DP4, and DP6), we received explicit comments \revise{that SyncUp could support}{about how SyncUp supported} the identification of ill-synchronized segments more accurately than a standard video player interface.
One reason mentioned by our participants is that people may have biases when they review practice videos.

\myquote{``If a person dances very well, then you will unconsciously focus your attention on that person, and ignore the others.
When you look back again, you may find they are synchronized quite well, but in fact, they are not.''}{DP4}

\myquote{``Usually there are lots of people dancing together, and I might not be able to notice whether they are synchronized or not.'' }{DL1}

They appreciated that our system could avoid \revise{such}{these} situations by \textit{``treating everyone equally''} (DP4).
Our participants also appreciated that SyncUp \revise{also }{}considers temporal alignments.
This feature \revise{can be}{is} helpful for those who are not well-trained \revise{for the}{with a} sense of tempo.

\myquote{``In fact, there truly exist people who are not sensitive to the tempo ... If there exists an app that can help this, it can effectively help those people, like me.''}{DP4}

One interesting anecdote was that an objective assessment offered by SyncUp can be more socially acceptable and could encourage communication among dancers.
\myquote{
``[Other] dancers are usually my friends ... If you directly say you are wrong here and there, it would hurt our friendship ... In the past, we actually noticed lots of places that we all were wrong. Because we were shy, we just kept silent.''}{DL3}

\myquote{
``It (SyncUp) is objective. For those highlighted places (segments), we can click one, and check it together. This encourages some proper communication.''
}{DL3}

\subsubsection{Enabling a Quick Iteration of Review and Practice}
Our participants confirmed the benefits of video recordings for their practices.
However, they also explicitly mentioned that they were \revise{not able}{unable} to review their videos at the place due to time constraints imposed by external factors:

\myquote{
``Mostly we watch these videos after we go back home because it is time-consuming. Also, there is a time limit in renting a dance studio.''}{DP4}

\myquote{
``Although we might occasionally watch [practice videos] on the spot, because we pay for renting the dance studio and the time is limited, it is impossible to examine videos in detail.''}{DP3}

Participants agreed that SyncUp could help them quickly review practice videos without viewing the full footage closely.
In particular, they appreciated the \textit{``high efficiency''} (DL1, DL2, DL3, DP1, DP2, DP3, and DP6) of our system. 
\myquote{
``If I use the system, it will automatically recognize those ill-synchronized segments. In this way, we can focus on these parts [during the practice], saving lots of time.''
}{DL2}
\myquote{
``The system can save time in many aspects of the practice. For example, we can record a video, and the system outputs the analysis, showing several segments. Then we can just quickly practice these parts (segments). Compared with slowly watching these videos and checking which parts (segments) we can improve after we go home, this (SyncUp) is very efficient.''
}{DP6}

\subsubsection{Other Potential Use of SyncUp}
\revise{Our participants}{Participants} also shared their potential creative use\revise{}{s} of SyncUp for their practices. 
DP6 explained their thought\revise{}{s} on how differently SyncUp could be useful depending on the dancer's skill level\revise{s}{}.  
\myquote{``"For those dancers who are already familiar with their dance routine, they would use it at a later stage. They could study some tiny issues of the pose similarity. However, for beginners, they might not be able to reach that stage, and would probably use this to help evaluate their tempo. Maybe from the early stage.''}{DP6} 

DP4 expressed their interests in using SyncUp to appreciate their good performances.
% \anran{}{I prefer appreciate, or we can say 'appreciate their good performances for self satisfaction.'}
They further mentioned that automatic extraction of \revise{such }{}good performance segments \revise{could}{would} be useful for sharing \revise{on}{with} online social networks.
\myquote{
``We record videos not only to find where we are out-of-synchronization, but also want to enjoy watching the movements that we are perfectly synchronized ... If this system can have these moments there (in the spotlight view), we can just download these videos and share them in our social app. We do not need to do video editing any more.''}{DP4}

\subsection{Discussion}
Our qualitative study with target users confirmed that SyncUp has a strong potential to support their practices.
In particular, they agreed that it would allow for a quick iterative practice process, which aligns with our goal.
Our results also imply that one major advantage of SyncUp is \revise{to offer an}{its} objective assessment \revise{on}{of} dance performances.
As DL3 commented, direct critical feedback on others' mistakes can be strongly discouraging for people \revise{with certain}{from certain} cultural backgrounds.
A system like SyncUp could serve as a mediator in such cases, and help group dancers maintain their relationships while allowing them to exchange honest feedback on their dance performances.

Our participants suggested several improvements to make SyncUp more practical.
These suggestions included a customizable synchronization score calculation method to reflect their preferences; an intelligent segmentation method that considers thematic changes of dances; and customizable discretization in overlay coloring (e.g., showing overlays only when the synchronization score becomes below a pre-defined threshold).

%% file: contents/conclusion.tex
\section{Limitations}

This work has limitations \revise{to be explicitly mentioned to clarify the contributions of this work.}{that we mentioned herein.}
The current interface design and functionalit\revise{ies}{y} are tailored toward synchronized dancing, choreography consisting of a series of synchronized poses and/or temporally-aligned movements among multiple dancers.
Although some of \revise{them}{the features} may be useful for other kinds of dancing or motions executed by multiple people, future work should examine whether such extensions would match \revise{users'}{user} requirements \revise{and desires}{}. 
For example, a Mexican wave (or a stadium wave) is \revise{one}{a} type of collective motions by multiple people which is beyond the scope of the current SyncUp implementation.
Supporting such a metachronal rhythm \revise{could}{would} be \revise{one}{an} interesting direction for future work.

In our system evaluations, our data only covered particular types of synchronized dancing (\revise{}{i.e., }dances originally performed by Asian pop stars or anime characters).
Although our performance quantification methods do not assume domain knowledge of dances, future work should study how applicable they would be to different types of synchronized dancing .

In our current implementation, the system uses 2D human skeletons.
Recent technology~\cite{muhammed2019vibe} can achieve 3D skeletons with a normal RGB camera, but the computation cost \revise{becomes}{is} much higher.
Our main contributions lie in \revise{integrating}{the intergration of} computer vision technology to enable \revise{a }{}quick review\revise{}{s} of synchronized dancing practice videos rather than improving the accuracy of pose similarity analysis in synchronized dancing.

Our temporal alignment analysis assumes that ground truth movements (i.e., dances by the leader) are given.
Another approach is to identify the beats of the background music and quantify how well dancers' motions are aligned with them.
We experimented with this, but decided not to incorporate it into the current SyncUp implementation.
We observed that dancers' movements are not always on the beats of the background music even though their dancing can be considered well synchronized.
This issue can be intrinsically challenging, and further explorations on human perception of temporal alignment in dances are encouraged.

In our current implementation, video analysis is performed \revise{in}{using} a remote computing resource in a post-hoc manner (i.e., after one practice session \revise{is over.}{was complete.}) instead of on a mobile device in real time. 
Thus, users would have a small amount of waiting time.
However, we do envision SyncUp would be able to \revise{offer}{provide} feedback while dancers \revise{would be}{are} taking a short break after running multiple practice sessions.
In this manner, SyncUp can contribute to mitigating large isolations between practices and reviews with recordings that dancers currently experience.
Future implementations may include more computationally\revise{-}{ }efficient pose detection and temporal alignment analysis, which may be executable for a smaller amount of time \revise{}{and/}or on a mobile device.
Recent computer vision technology has achieved real-time human pose detection on mobile devices~\cite{Lugaresi2019MediaPipe}\revise{though}{, however,} the extension to multi-person pose detection is still under development.
We note that our scope of this work is the development of dance analysis methods tailored toward synchronized dancing and both system and user evaluations on the proposed system instead of improving the computational costs of the analysis process.
Our work encourages further research and development of such computer vision technolog\revise{y}{ies} by demonstrating application\revise{}{s} and interface design\revise{}{s} that fit\revise{s}{} users' needs and workflow.

\section{Conclusion}
Synchronized dancing \revise{is attracting interests by}{attracts} amateur dancers, but interactive support for its practice is still insufficient.
This paper presents SyncUp, a vision-based interactive system \revise{including}{that includes} multiple visualizations and \revise{offering}{offers} quick access to segments where dances are not well synchronized.
The system integrates two approaches to quantify the performance of synchronized dancing: pose similarity and temporal alignment.
Our system evaluations confirm that performance predictions by SyncUp are highly correlated with human ratings.
Participants in our qualitative user study shared their positive opinions on the features in SyncUp, and \revise{also}{they} commented several potential use\revise{}{s} of the system.
Future work should conduct an in-depth user study through the deployment of SyncUp into actual practice\revise{s}{ sessions} to validate the results reported in this paper.

\section*{Acknowledgments}
We would like to thank Arissa J. Sato, Carla F. Griggio, Zefan Sramek, and all the anonymous reviewers for their valuable feedback on our manuscript. 
We also appreciate Minghui Chen for his support to create the demonstration video of this project.
We finally want to show our gratitude to all participants in our user studies, interviews, and surveys for their invaluable comments that are inspiring for both our work and future work.
This research received support from the NII CRIS collaborative research program jointly managed by NII CRIS and LINE Corporation.
% Carla, Arissa, Zefan for proofreading
% Iraka for video
% CRIS funding

%% file: contents/appendix.tex
%TC:ignore 
\appendix

\vspace{1cm}

\section{Examples of Pose Similarity Visualization}
\begin{figure}[H]
    \begin{subfigure}[b]{0.55\textwidth}  
        \centering 
        \includegraphics[width=\textwidth]{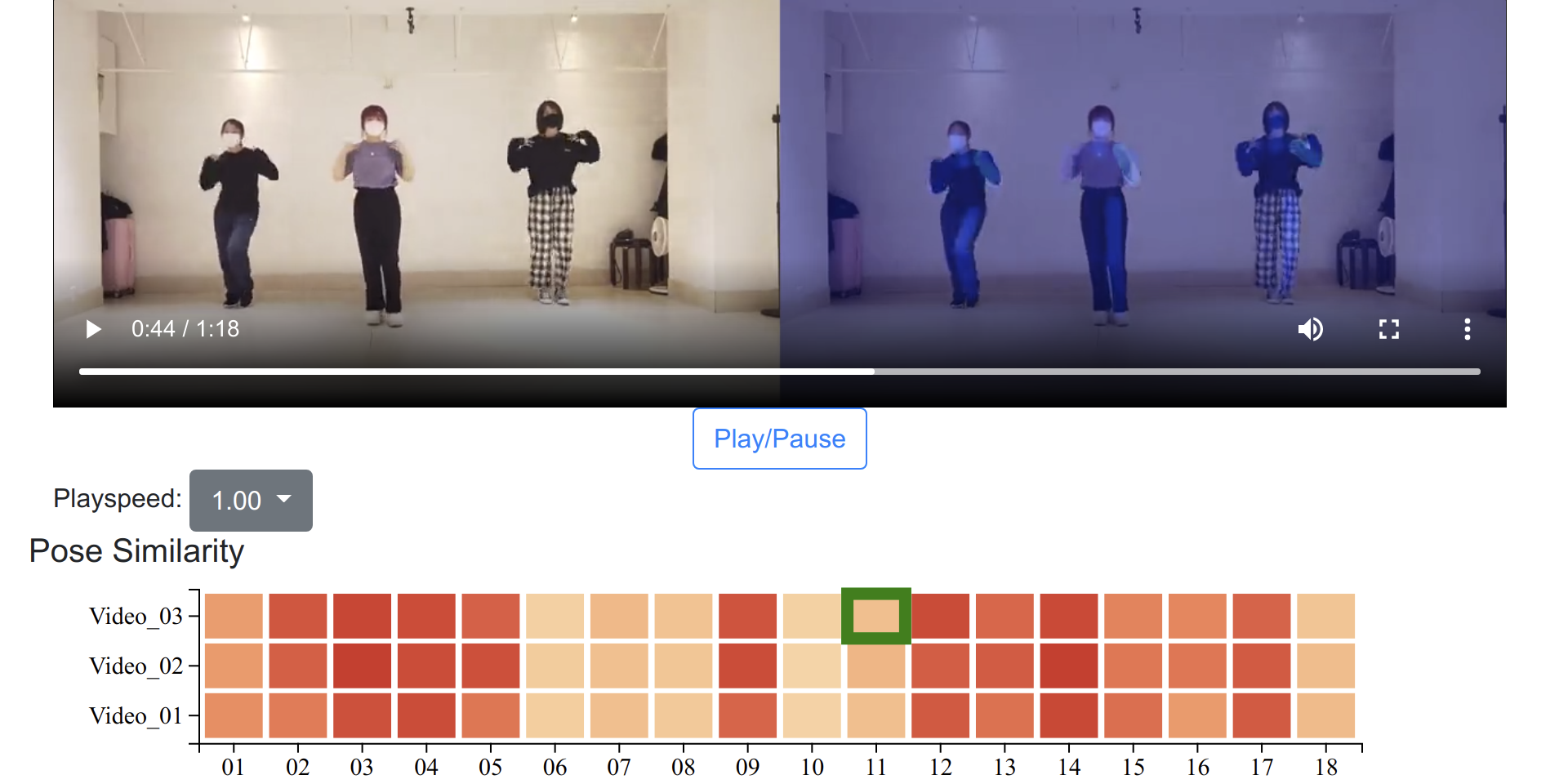}
        \caption[]%
        {With a high pose similarity score}    
        \label{fig:yellow}
    \end{subfigure}
    \begin{subfigure}[b]{0.55\textwidth}  
        \centering 
        \includegraphics[width=\textwidth]{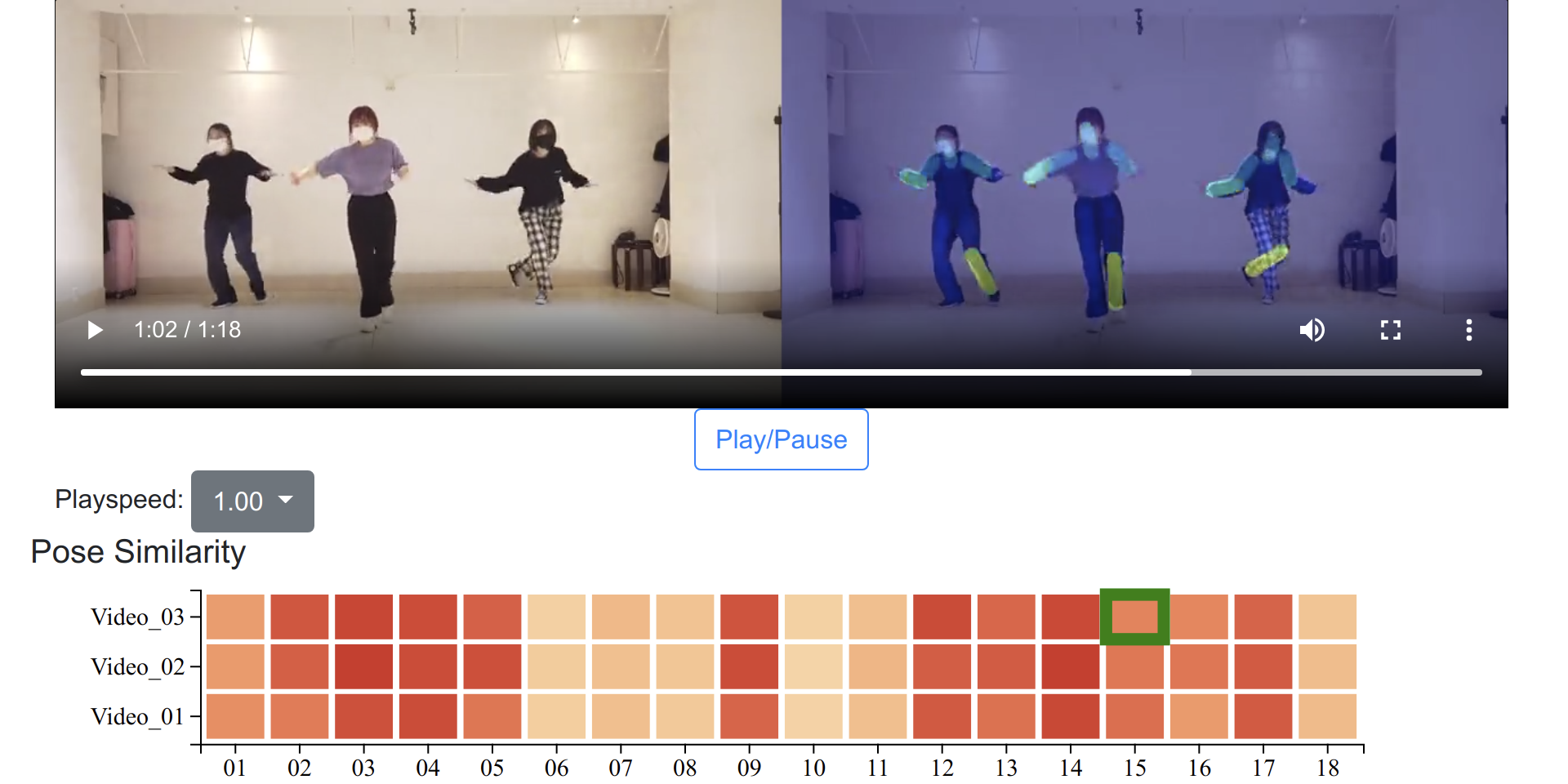}
        \caption[]%
        {With a middle pose similarity score}    
        \label{fig:orange}
    \end{subfigure}
    \begin{subfigure}[b]{0.55\textwidth}  
        \centering 
        \includegraphics[width=\textwidth]{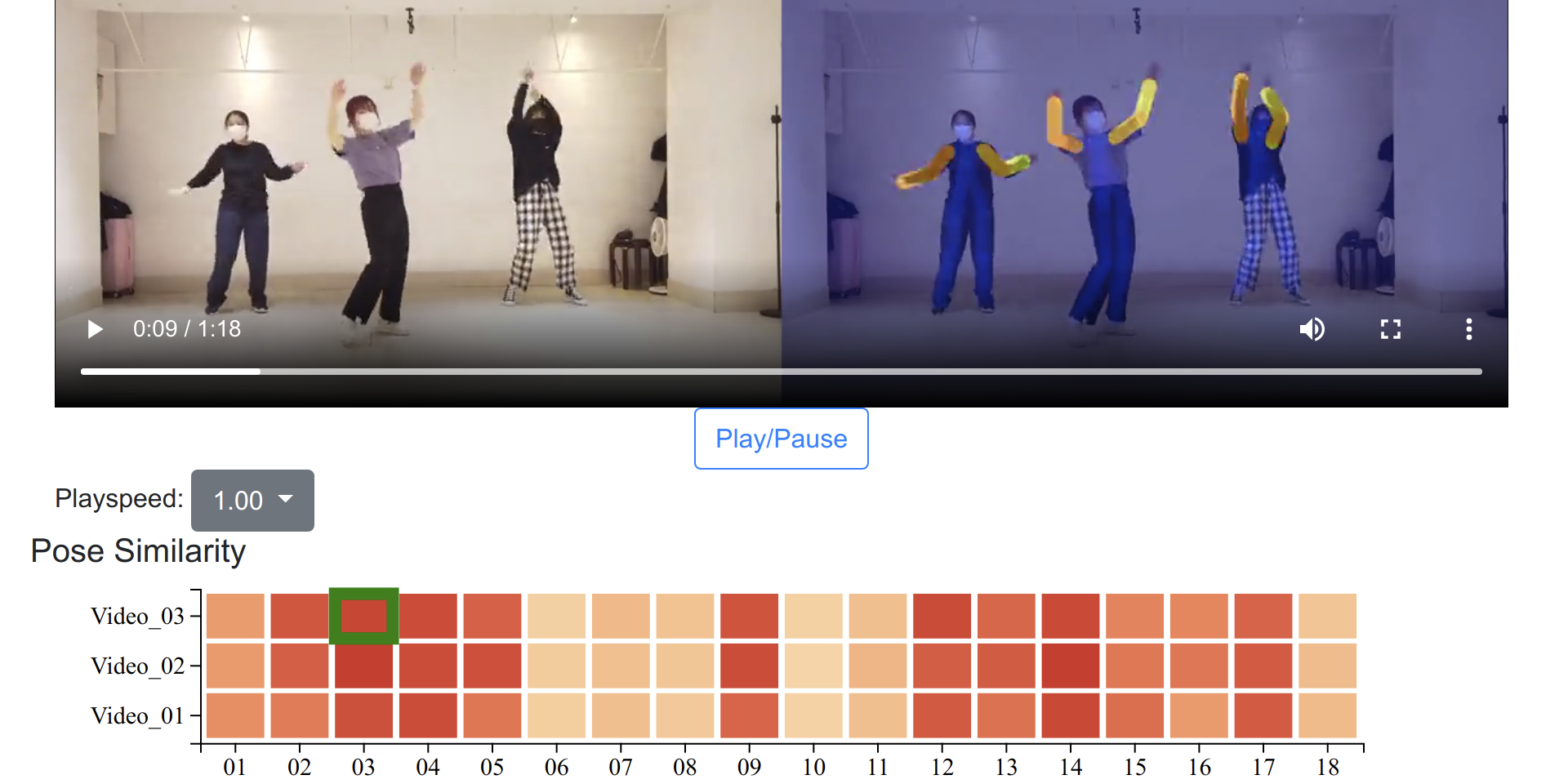}
        \caption[]%
        {With a low pose similarity score}    
        \label{fig:red}
    \end{subfigure}
    \caption{Three examples of visualization related to pose similarity. These\revise{ three}{} examples were chosen to illustrate \revise{the }{}cases where the pose similarity score was low, in the middle, and high.}
    \label{fig:appendix_pose_sim_vis}
\end{figure}

Figure~\ref{fig:appendix_pose_sim_vis} provides three examples of side-by-side views with heatmap overlays along\revise{ with for the}{side} pose similarity score visualization\revise{}{s}.
\revise{We n}{N}ote that a darker red color in the 1D heatmap visualization indicates a lower pose similarity score.

\clearpage

\section{Example frames with high prediction errors in the analysis of pose similarity}
\label{appendix}
\begin{figure}[H]
    \begin{subfigure}[b]{0.24\textwidth}  
        \centering 
        \includegraphics[width=\textwidth]{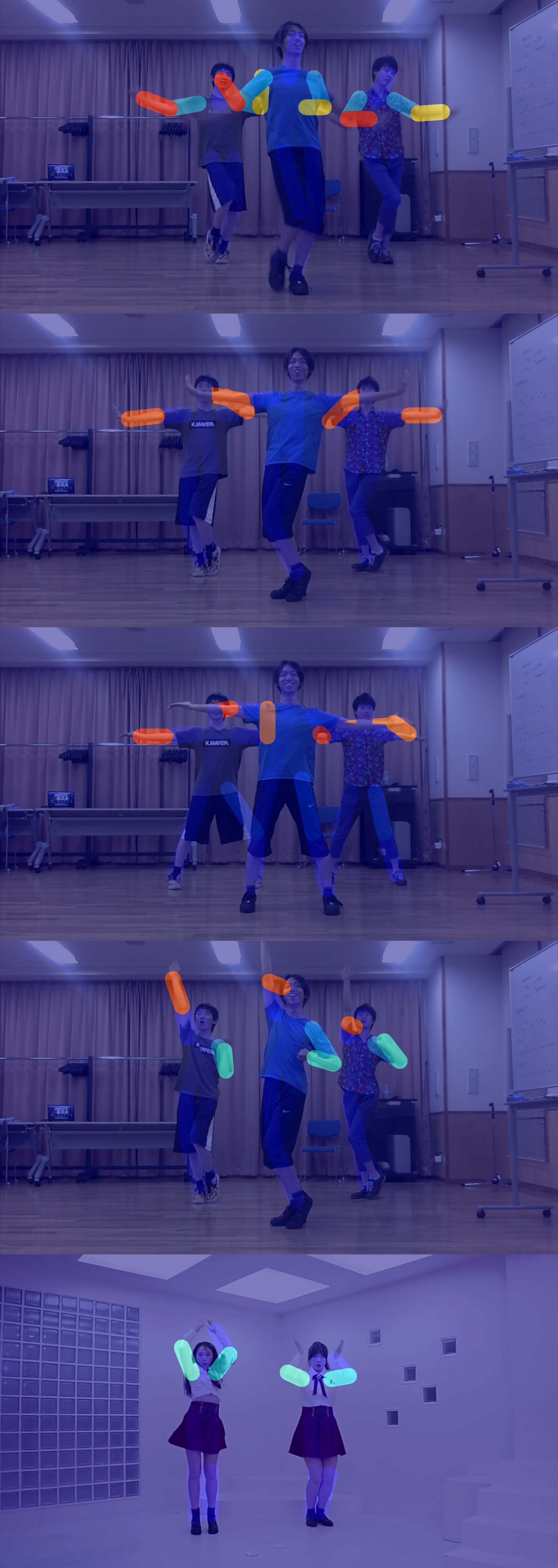}
        \caption[]%
        {{\small SVR}}    
        \label{fig:pose_sim:svm}
    \end{subfigure}
    % \quad
    \begin{subfigure}[b]{0.24\textwidth}
        \centering
        \includegraphics[width=\textwidth]{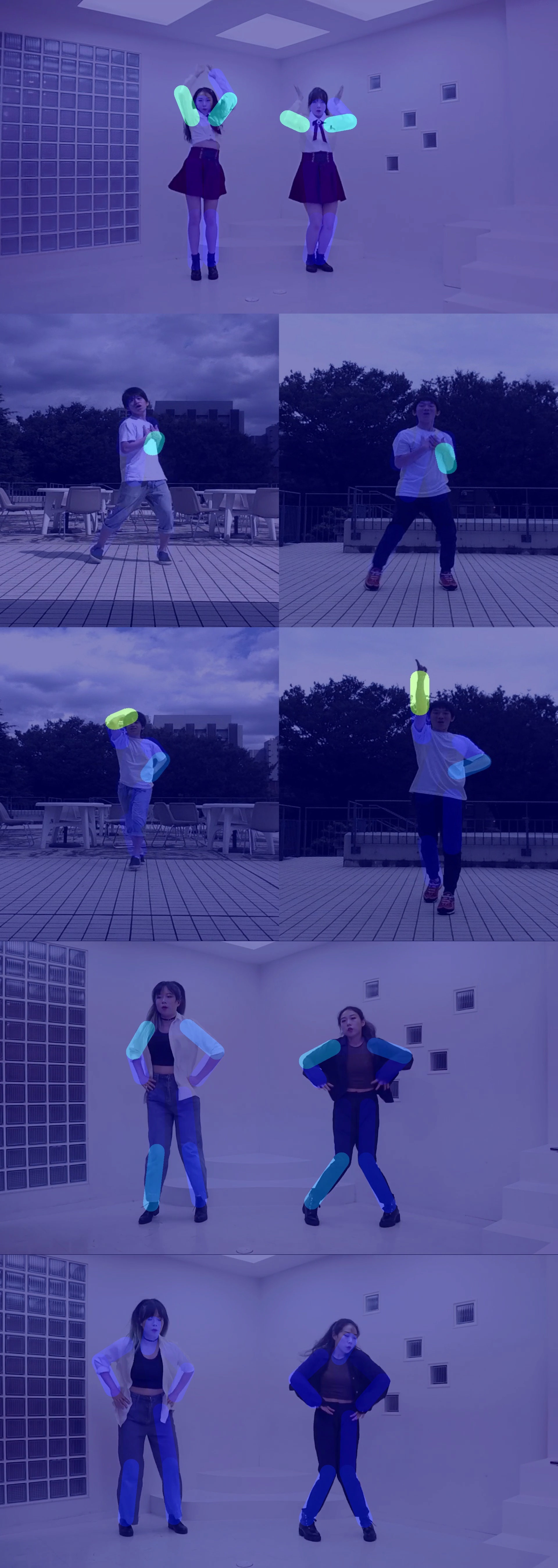}
        \caption[Network2]%
        {{\small Simple Addition}}    
        \label{fig:pose_sim:add}
    \end{subfigure}
    \begin{subfigure}[b]{0.24\textwidth}   
        \centering 
        \includegraphics[width=\textwidth]{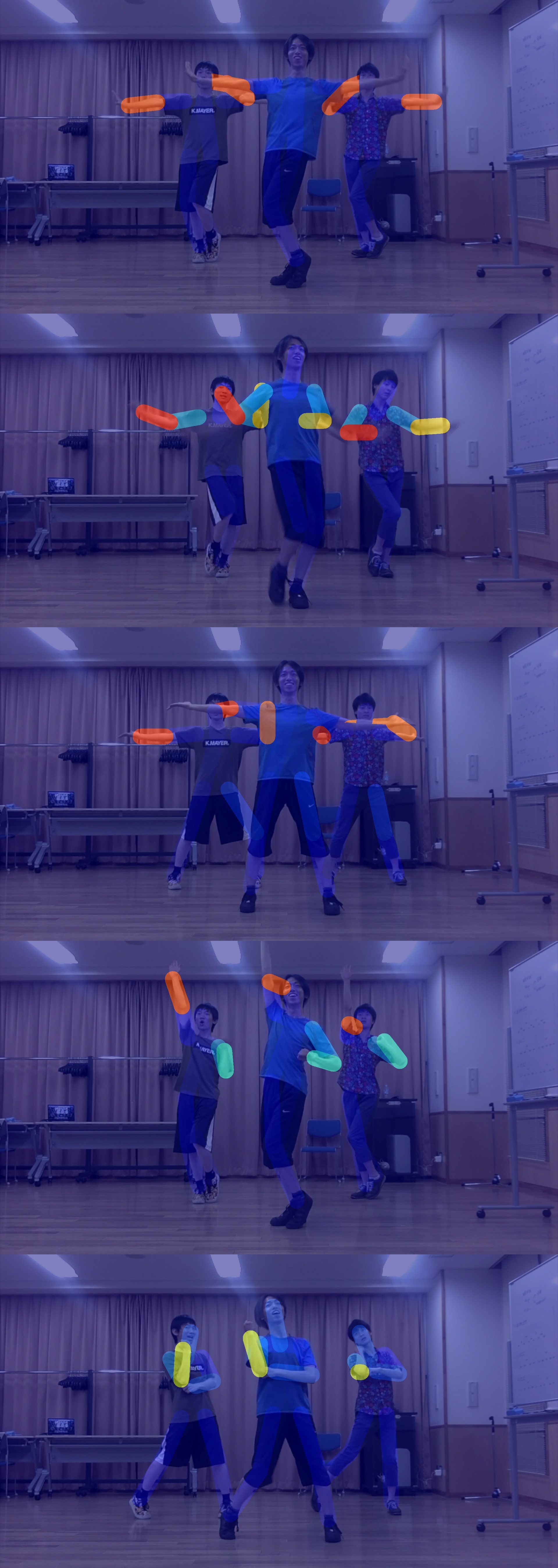}
        \caption[]%
        {{\small NN\_short}}    
        \label{fig:pose_sim:nn_short}
    \end{subfigure}
    \begin{subfigure}[b]{0.24\textwidth}   
        \centering 
        \includegraphics[width=\textwidth]{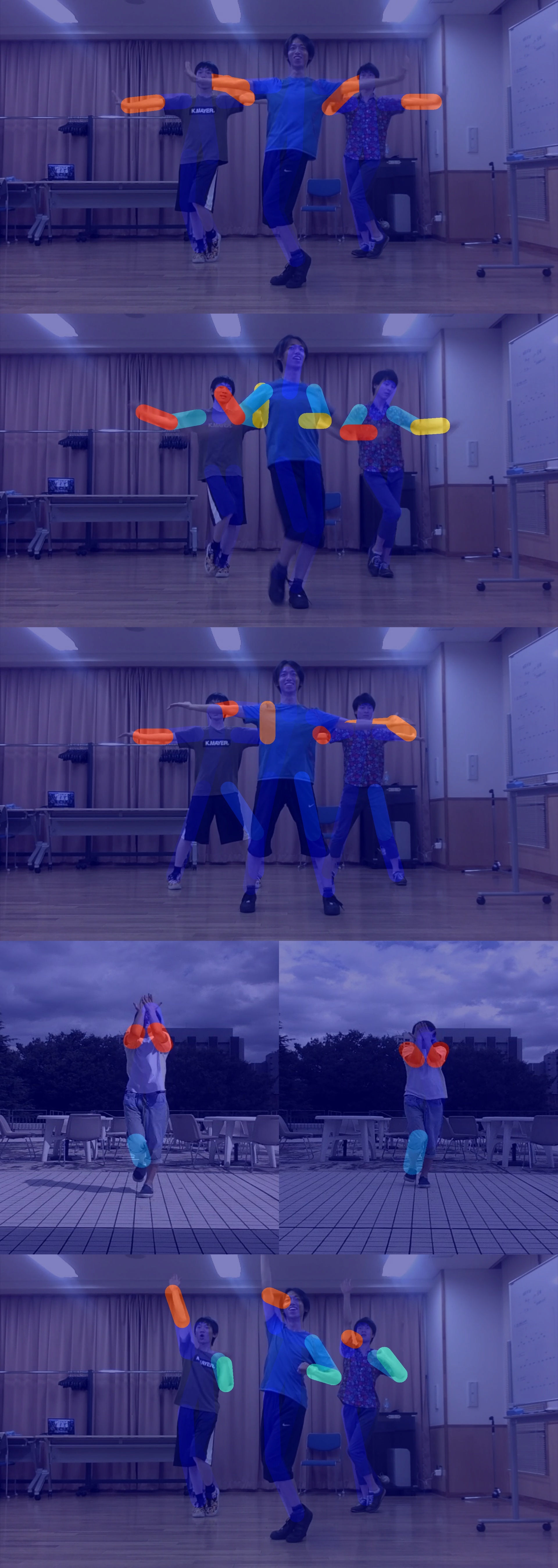}
        \caption[]%
        {{\small NN\_long}}    
        \label{fig:pose_sim_eval:nn_long}
    \end{subfigure}
    \caption
    {\small \revise{The f}{F}rames that caused the five largest pose similarity prediction errors with each OPS quantification method.
    The top figure in each column shows the frame that exhibited the largest error with the corresponding method.} 
    \label{fig:topk}
\end{figure}

Figure~\ref{fig:topk} shows frames that caused the five largest pose similarity prediction errors with each OPS quantification method.
The top figure in each column shows the frame that exhibited the largest error with the corresponding method.
This result suggests that one major cause of large pose similarity prediction error\revise{s}{} was occlusion of body parts.
Other \revise{}{error }attributes \revise{of the errors }{}include lack of hand pose detection (the bottom \revise{figure in}{of} Figure~ \ref{fig:pose_sim:svm}) and lack of 3D pose detection support\revise{s}{} (the fourth figure in Figure~\ref{fig:pose_sim_eval:nn_long}).
Different from our learning-based \revise{}{method }in which errors are caused by our dependent pose detection algorithms, we observed that the Simple Addition method makes high-error predictions even when the pose detection is correct. 
Such hard-to-interpret errors \revise{would}{may} confuse users and developers in practice.

